\documentclass[11pt]{article}
\usepackage{adjustbox}
\usepackage{mathptmx}
\usepackage{rotating}
\usepackage{bm}
\usepackage{amsfonts}
\usepackage[colorlinks,citecolor=blue,urlcolor=blue,hyperfootnotes=false]{hyperref}
\usepackage{amssymb}
\usepackage[bottom, multiple]{footmisc}
\usepackage[hyperref]{ntheorem}
\usepackage{natbib}
\usepackage{newtxtext}
\usepackage{graphicx}
\usepackage{float}
\usepackage[doublespacing,nodisplayskipstretch]{setspace}
\usepackage[top=1in, bottom=1in, left=1in, right=1in]{geometry}
\usepackage{caption}
\usepackage[margin=18pt, font+=small, labelformat=parens,labelsep=space, skip=6pt, list=false,hypcap=false]{subcaption}
\usepackage{booktabs}
\usepackage{threeparttablex}
\usepackage{threeparttable}
\usepackage{amsmath}
\usepackage{lscape}
\usepackage{scrextend}
\usepackage{relsize}
\usepackage{multicol}
\usepackage{chngcntr}
\usepackage{etoolbox}
\usepackage{amssymb}
\usepackage{tabularx}
\usepackage[english]{babel}
\usepackage{multirow}
\usepackage{booktabs}
\usepackage[labelfont=bf]{caption}
\usepackage{float}
\usepackage{titlesec}
\usepackage{array}
\usepackage{color}
\usepackage{afterpage}
\usepackage{framed}
\usepackage{float}
\usepackage[titletoc,title]{appendix}
\usepackage{enumerate}
\usepackage{longtable}
\usepackage{microtype}
\usepackage{hyphenat}
\usepackage{autobreak}
\usepackage{xr}

\numberwithin{equation}{section}
\newdimen \dummy
\dummy=\oddsidemargin
\titlespacing \section{0pt}{1.0ex plus -1ex minus -.2ex}{-\parskip}
\titlespacing \subsection{0pt}{1.0ex plus -1ex minus -.2ex}{-\parskip}
\AtBeginDocument{
\addtolength{\abovedisplayskip}{-1ex}
\addtolength{\belowdisplayskip}{-1ex}
}
\newcolumntype{x}[1]{>{\centering \arraybackslash \hspace{0pt}}p{#1}}
\allowdisplaybreaks
\providecommand{\U}[1]{\protect\rule{.1in}{.1in}}
\theoremstyle{plain}
\newtheorem{theorem}{Theorem}
\newtheorem{corollary}{Corollary}

\newtheorem{proposition}{Proposition}

\newtheorem{assumption}{Assumption}

\theorembodyfont{\normalfont}
\newtheorem{remark}{Remark}
\deffootnote{1em}{1.6em}{\thefootnotemark \enskip}

\input{tcilatex}
\begin{document}

\title{Doubly Robust Difference-in-Differences Estimators\thanks{%
First complete version: November 29, 2018. We thank the editor, Serena Ng, the
associate editor, two anonymous referees, Brantly Callaway, Alex Poirier,
Vitor Possebom, Yuya Sasaki, Tymon S\l oczy\'{n}ski, Qi Xu, and the
audiences of the 2018 SEA conference, 2019 New York Econometrics Camp, and
the 2019 IAAE conference for valuable comments.}}
\author{Pedro H. C. Sant'Anna\thanks{%
Department of Economics, Vanderbilt University. E-mail:
pedro.h.santanna@vanderbilt.edu. Part of this article was written when I was
visiting the Cowles Foundation at Yale University, whose hospitality is
gratefully acknowledged.} \\
%EndAName
Vanderbilt University \and Jun B. Zhao\thanks{%
Department of Economics, Vanderbilt University. E-mail:
jun.zhao@vanderbilt.edu.} \\
%EndAName
Vanderbilt University}
\date{May 5, 2020}
\maketitle

\begin{abstract}
\onehalfspacing{
This article proposes doubly robust estimators for the average treatment effect on the treated (ATT) in 
difference-in-differences (DID) research designs.  
In contrast  to alternative DID estimators, the proposed estimators are consistent if either
 (but not  necessarily both) a propensity score or outcome regression working models are correctly specified.
  We also derive the semiparametric efficiency bound for the ATT in DID designs when either panel or repeated 
cross-section data are available, and show that our proposed estimators attain the semiparametric efficiency
 bound when the working models  are correctly specified. Furthermore, we quantify the potential efficiency gains of having access to panel data instead of repeated cross-section data.  Finally, by paying particular attention to the estimation
 method used to estimate the nuisance parameters, we show that one can sometimes construct doubly robust DID 
estimators for the ATT that are also doubly robust for inference. Simulation studies and an 
empirical application illustrate the desirable finite-sample performance of the proposed estimators. 
Open-source software for implementing the proposed policy evaluation tools is available.
}
\end{abstract}

\setlength{\abovedisplayskip}{3pt} \setlength{\belowdisplayskip}{3pt}

\pagebreak

\section{Introduction\label{sec:intro}}

Difference-in-differences (DID) methods are among the most popular
procedures practitioners adopted to conduct policy evaluation with
observational data. In its canonical form, DID identifies the average
treatment effect on the treated (ATT) by comparing the difference in pre and
post-treatment outcomes of two groups: one that receives and one that does
not receive the treatment (the treated and comparison group, respectively).
In order to attach a causal interpretation to DID estimators, researchers
routinely invoke the (unconditional) parallel trends assumption (PTA): in
the absence of the treatment, the average outcome for the treatment and
comparison groups would have followed parallel paths over time. Although the
PTA is fundamentally untestable, its plausibility is usually questioned if
the observed characteristics that are thought to be associated with the
evolution of the outcome are not balanced between the treated and comparison
group. In such cases, researchers usually deviate from the canonical DID
setup and incorporate pre-treatment covariates into the DID analysis and
assume that the PTA is satisfied only after conditioning on these covariates.

In this paper, we study the robustness and efficiency properties of DID
estimators for the ATT when the PTA holds after conditioning on covariates.
We consider both settings where panel data are available and settings where
only repeated cross-section data are available. We contribute to the DID
literature in different fronts. First, we derive doubly robust (DR)
estimands for the ATT under DID settings and propose DR DID estimators for
the ATT that are consistent when either a working (parametric) model for the
propensity score or a working (parametric) model for the outcome evolution
for the comparison group is correctly specified. The setting where only
repeated cross-section data are available is particularly interesting. We
propose two different DR DID estimators for the ATT that differ from each
other depending on whether or not one models the outcome regression for the
treated group in both pre and post-treatment periods. Nonetheless, we show
DR property does not depend on such a choice.

Second, we derive the semiparametric efficiency bounds for the ATT under DID
designs. The semiparametric efficiency bounds we derive are nonparametric in
the sense that we do not assume researchers have additional knowledge about
outcome regressions or the propensity score functional forms. As so, these
bounds provide a standard against which one can compare the efficiency of
any (regular) semiparametric DID estimator for the ATT. Here, it is also
worth stressing that these semiparametric efficiency bounds explicitly
incorporate all the restrictions implied by the invoked identification
assumptions. Importantly, these restrictions differ depending on whether
panel or repeated cross-section data are available. In both cases they
involve the moment restrictions implied by the conditional PTA, though, when
repeated cross-section data are available, they also include the
restrictions implied by the identifying assumption that the joint
distribution of covariates and treatment status is invariant to the sampling
period (pre and post-treatment). We emphasize that failing to account for
all these implied restrictions can lead to discrepancies on the derived
efficiency bound, which, in turn, may suggest that some estimator is
semiparametrically efficient when, in fact, it is not.

With the semiparametric efficiency bounds at hand, we can answer several
questions that one may have. For instance, one may wonder whether there are
efficiency gains associated with having access to panel instead of repeated
cross-section data. By directly comparing the efficiency bounds under these
two setups, we not only show that the answer to the aforementioned question
is yes, but also show that such gains tend to be larger when the sample
sizes of the pre and post-treatment repeated cross-section data are more
imbalanced.

Another natural question that arises is whether our proposed DR DID
estimators can attain the semiparametric efficiency bound. We show that when
the working models for the propensity score and for the outcome evolution
for the comparison group are correctly specified, our proposed DR DID
estimator for the panel data setup is locally efficient, though the DR DID
estimators for the cross-section setup are not. In fact, when only repeated
cross-section data are available, we show that our proposed DR DID estimator
that relies on modelling the propensity score and the outcome evaluation of 
\emph{both} the treated and comparison groups attains the semiparametric
efficiency bound when all working models are correctly specified. We
quantify the loss of efficiency associated with using the inefficient\ DR
DID estimator instead of the locally efficient one, and illustrate via Monte
Carlo simulations that such a loss can indeed be large.

Our proposed methodology accommodates linear and nonlinear working models
for the nuisance functions. We establish $\sqrt{n}$-consistency and
asymptotic normality of the proposed DR DID estimators when generic
parametric working models are used for the nuisance functions. In doing so
we emphasize that, in general, the DR property of our estimators is with
respect to consistency and not to inference. In other words, the exact form
of the asymptotic variance of our proposed estimators depends on whether the
propensity score and/or the outcome regression models are correctly
specified. Given that, in practice, one does not know a priori which models
are correctly specified, one should consider the estimation effects from all
first-step estimators when estimating the asymptotic variance. Failing to do
so may lead to invalid inference procedures.

Motivated by this observation, a third contribution of this paper is to show
that, by paying particular attention to the estimation method used for
estimating the nuisance parameters, it is sometimes possible to construct
computationally simple DID estimators for the ATT that are not only DR
consistent and locally semiparametric efficient, but are also doubly robust
for inference. These further improved DR DID estimators are particularly
attractive and easy to implement when researchers are comfortable with a
logistic working model for the propensity score and with linear regression
working models for the outcome of interest.

\textbf{Related literature:} Our proposal builds on two branches of the
causal inference literature. First, our methodological results are
intrinsically related to other DID papers; for an overview, see e.g.,
Section 6.5 of \cite{Imbens2009} and references therein. Two leading
contributions in this branch of literature that are particularly relevant to
this paper are \cite{Heckman1997}, who propose kernel-based DID regression
estimators, and \cite{Abadie2005}, who proposes (parametric and
nonparametric) DID inverse probability weighted (IPW) estimators. We note
that when the dimension of available covariates is high or even moderate,
fully nonparametric procedures usually do not lead to informative inference
because of the \textquotedblleft curse of dimensionality\textquotedblright .
In these cases, researchers often adopt parametric methods. Our DR DID
estimators fall in this latter category.

Second, our results are also directly related to the literature on doubly
robust estimators, see \cite{Robins1994}, \cite{Scharfstein1999}, \cite%
{Bang2005}, \cite{Wooldridge2007a}, \cite{Chen2008}, \cite{Cattaneo2010}, 
\cite{Graham2012, Graham2016}, \cite{Vermeulen2015}, \cite{Lee2017}, \cite%
{Sloczynski2018}, \cite{Rothe2018}, \cite{Muris2019}, among many others; for
an overview, see section 2 of \cite{Sloczynski2018}, and \cite{Seaman2018}.
Recently, DR estimators have also been playing an important role when one
uses data-adaptive, \textquotedblleft machine learning\textquotedblright\
estimators for the nuisance functions, see e.g., \cite{Belloni2014}, \cite%
{Farrell2015}, \cite{Chernozhukov2017}, \cite{Belloni2017}, and \cite%
{Tan2019}. As so, these papers are also broadly related to our proposal,
even though we use parametric first-step estimators. On the other hand, we
note that the aforementioned papers focus on either the \textquotedblleft
selection on observables\textquotedblright\ or \textquotedblleft
IV/LATE\textquotedblright\ type assumptions, whereas we pay particular
attention to the conditional DID design. Thus, our results complement theirs.

To derive the semiparametric efficiency bounds for the ATT under\ the DID
framework, we build on \cite{Hahn1998} and \cite{Chen2008}. Although we
follow the structure of semiparametric efficiency bound derivation of the
aforementioned papers (which, in turn, follow \cite{Newey1990}), our derived
semiparametric efficiency bounds complement theirs as we focus on DID
designs while \cite{Hahn1998} and \cite{Chen2008} results rely on
\textquotedblleft selection on observables\textquotedblright\ type
assumptions in cross-section setups.

Our results for the further improved DR DID estimators build on \cite%
{Vermeulen2015}, who propose estimators that are DR for inference in
cross-section setups under selection on observables type assumptions. We
extend \cite{Vermeulen2015} proposal to DID settings with both panel and
repeated cross-section data. Our further improved DR DID estimators also
build on \cite{Graham2012}, as their proposed propensity score estimator is
one important component of our proposal.

Finally, in work related but independent from ours, \cite{Zimmert2019}
provides high-level conditions under which one can use \textquotedblleft
machine-learning\textquotedblright\ first-step estimators when estimating
the ATT in DID setups. His results complement ours, though we note that his
proposed estimators for the repeated cross-section case do not attain the
semiparametric efficiency bound derived in this paper, and the loss of
efficiency can be of first-order importance. We also note that \cite%
{Zimmert2019} does not provide a detailed comparison between the panel and
repeated cross-section data setups like we do, nor discusses DR inference
procedures, which are particularly relevant under model misspecifications.

\textbf{Organization of the paper:} In the next section, we describe this
paper's framework, briefly give an overview of the existing DID estimators
and describe how we combine the strengths of each method to form our DR DID
estimands. We also derive semiparametric efficiency bounds for the ATT in
Section \ref{sec:did}. In Section \ref{sec:asy.theory}, we propose different
DR DID estimators, derive their large sample properties, and show that we
can get improved DR DID estimators by paying particular attention to the
estimation method used for estimating the nuisance parameters. We examine
the finite sample properties of our proposed methodology by means of a Monte
Carlo study in Section \ref{sec:MC}, and provide an empirical illustration
in Section \ref{sec:applications}. Section \ref{sec:conclusion} concludes.
Mathematical proofs are gathered in the Supplemental Appendix.\footnote{%
The Supplemental Appendix is available at %
\url{https://pedrohcgs.github.io/files/DR-DID Appendix.pdf}} Finally, all
proposed policy evaluation tools discussed in this article can be
implemented via the open-source R package \texttt{DRDID}, which is freely
available from GitHub (https://github.com/pedrohcgs/DRDID).

\section{Difference-in-differences\label{sec:did}}

\subsection{Background \label{sec:framework}}

We first introduce the notation we use throughout the article. We focus on
the case where there are two treatment periods and two treatment groups. Let 
$Y_{it}$ be the outcome of interest for unit $i$ at time $t$. We assume that
researchers have access to outcome data in a pre-treatment period $t=0$ and
in a post-treatment period $t=1$. Let $D_{it}=1$ if unit $i$ is treated
before time $t$ and $D_{it}=0$ otherwise. Note that $D_{i0}=0$ for every $i$%
, allowing us to write $D_{i}=D_{i1}$. Using the potential outcome notation,
denote $Y_{it}\left( 0\right) $ the outcome of unit $i$ at time $t$ if it
does not receive treatment by time $t$ and $Y_{it}\left( 1\right) $ the
outcome for the same unit if it receives treatment. Thus, the realized
outcome for unit $i$ at time $t$ is $Y_{it}=D_{i}Y_{it}\left( 1\right)
+\left( 1-D_{i}\right) Y_{it}\left( 0\right) $. A vector of pre-treatment
covariates $X_{i}$ is also available. Henceforth, we assume that the first
element of $X_{i}$ is a constant.

In the rest of the article, we assume that either panel or repeated
cross-section data on $\left( Y_{it},D_{i},X_{i}\right) $, $t=0,1$ are
available. When repeated cross-section data are available, we follow \cite%
{Abadie2005} and assume that covariates and treatment status are stationary.
We formalize these conditions in the following assumption. Let $T_{i}$ be a
dummy variable that takes value one if the observation $i$ is only observed
in the post-treatment period, and zero if observation $i$ is only observed
in the pre-treatment period. Define $Y_{i}=T_{i}Y_{i1}+\left( 1-T_{i}\right)
Y_{i0}$, and let $n_{1}$ and $n_{0}$ be the sample sizes of the
post-treatment and pre-treatment periods such that $n=n_{1}+n_{0}$. Finally,
let $\lambda =\mathbb{P}\left( T=1\right) \in \left( 0,1\right) $.

\begin{assumption}
\label{ass:sampling}Assume that either $\left( a\right) $ the data $%
\{Y_{i0},Y_{i1},D_{i},X_{i}\}_{i=1}^{n}$ are independent and identically
distributed $(iid)$; or $\left( b\right) $ the pooled repeated cross-section
data $\left\{ Y_{i},D_{i},X_{i},T_{i}\right\} _{i=1}^{n}$ consist of $iid$
draws from the mixture distribution%
\begin{eqnarray*}
P\left( Y\leq y,D=d,X\leq x,T=t\right) &=&t\cdot \lambda \cdot P\left(
Y_{1}\leq y,D=d,X\leq x|T=1\right) \\
&&+\left( 1-t\right) \cdot \left( 1-\lambda \right) P\left( Y_{0}\leq
y,D=d,X\leq x|T=0\right) ,
\end{eqnarray*}%
where $\left( y,d,x,t\right) \in \mathbb{R\times }\left\{ 0,1\right\} \times 
\mathbb{R}^{k}\times \left\{ 0,1\right\} $, with the joint distribution of $%
\left( D,X\right) $ being invariant to $T$.
\end{assumption}

Assumption \ref{ass:sampling}$(a)$ covers the case where panel data are
available, whereas Assumption \ref{ass:sampling}$(b)$ covers the case where
repeated cross-section data are available, and allows for different sampling
schemes. For instance, it accommodates the binomial sampling scheme where an
observation $i$ is randomly drawn from either $\left( Y_{1},D,X\right) $ or $%
\left( Y_{0},D,X\right) $ with fixed probability $\lambda $ (here, $T$ is a
non-degenerated random variable). It also accommodates the \textquotedblleft
conditional\textquotedblright\ sampling scheme where $n_{1}$ observations
are sampled from $\left( Y_{1},D,X\right) $, $n_{0}$ observations are
sampled from $\left( Y_{0},D,X\right) $ and $\lambda =n_{1}/n$ (here, $T$ is
treated as fixed). On the other hand, Assumption \ref{ass:sampling}$\left(
b\right) $ rules out settings with compositional changes in $\left(
D,X\right) $, see e.g. \cite{Hong2013} for a discussion.

The parameter of interest is the average treatment effect on the treated, 
\begin{equation*}
\tau =\mathbb{E}[Y_{i1}(1)-Y_{i1}(0)|D_{i}=1].
\end{equation*}%
As expectations are linear operators and $Y_{i1}\left( 1\right) =Y_{i1}$ if $%
D_{i}=1$, we can rewrite the ATT as\footnote{%
Throughout the rest of the paper, to ease the notation burden we denote $%
\mathbb{E}\left[ \cdot \right] $ as generic expectations. In the case of
panel data, such expectations are with respect to the distribution of $%
\left( Y_{0},Y_{1},D,X\right) $. In the case of repeated cross-section data,
the expectations are with respect to the mixture distribution $\sum_{t=0}^{1}%
\mathbb{P}\left( T=t\right) \cdot \mathbb{P}\left( Y_{t}\leq y,D=d,X\leq
x|T=t\right) $.}%
\begin{equation}
\tau =\mathbb{E}[Y_{1}(1)|D=1]-\mathbb{E}[Y_{1}(0)|D=1]=\mathbb{E}%
[Y_{1}|D=1]-\mathbb{E}[Y_{1}(0)|D=1],  \label{ATT}
\end{equation}%
where we drop subscript $i$ to ease notation; we follow this convention
throughout the paper. From the above representation, it is clear that the
main challenge in identifying the ATT is to compute $\mathbb{E}%
[Y_{i1}(0)|D_{i}=1]$ from the observed data. To overcome this challenge, we
invoke the following assumptions.

\begin{assumption}
\label{ass:cond.pta} $\mathbb{E}[Y_{1}(0)-Y_{0}(0)|D=1,X]=\mathbb{E}%
[Y_{1}(0)-Y_{0}(0)|D=0,X]$ almost surely (a.s.).
\end{assumption}

\begin{assumption}
\label{ass:common.support} For some $\varepsilon >0$, $\mathbb{P}\left(
D=1\right) >\varepsilon $ and $\mathbb{P}\left( D=1|X\right) \leq
1-\varepsilon $ a.s..
\end{assumption}

Assumption \ref{ass:cond.pta}, which we refer to as the conditional PTA
throughout the paper, states that in the absence of treatment, the average
conditional outcome of the treated and the comparison groups would have
evolved in parallel. Note that Assumption \ref{ass:cond.pta} allows for
covariate-specific time trends, though it rules out unit specific trends.
Assumption \ref{ass:common.support} is an overlap condition and states that
at least a small fraction of the population is treated and that for every
value of the covariates $X$, there is at least a small probability that the
unit is not treated. These two assumptions are standard in conditional DID
methods, see e.g. \cite{Heckman1997}, \cite{Heckman1998}, \cite%
{Blundell2004a}, \cite{Abadie2005} and \cite{Bonhomme2011}.

Under Assumptions \ref{ass:sampling}-\ref{ass:common.support}, there are two
main flexible estimation procedures to estimate the ATT: the outcome
regression (OR) approach, see e.g. \cite{Heckman1997}, and the IPW approach,
see e.g. \cite{Abadie2005}. The OR approach relies on researchers ability to
model the outcome evolution. In such cases, under the aforementioned
assumptions one can estimate the ATT using%
\begin{equation}
\widehat{\tau }^{reg}=\bar{Y}_{1,1}-\left[ \bar{Y}_{1,0}+n_{treat}^{-1}%
\sum_{i|D_{i}=1}\left( \widehat{\mu }_{0,1}\left( X_{i}\right) -\widehat{\mu 
}_{0,0}\left( X_{i}\right) \right) \right] ,  \label{att.reg1}
\end{equation}%
where $\bar{Y}_{d,t}=\sum_{i\left\vert D_{i}=d,T_{i}=t\right.
}Y_{it}/n_{d,t} $ is the sample average outcome among units in treatment
group $d$ and time $t$, and $\widehat{\mu }_{d,t}\left( x\right) $ is an
estimator of the true, unknown $m_{d,t}\left( x\right) \equiv \mathbb{E}%
[Y_{t}|D=d,X=x]$,\footnote{%
In the repeated cross-section case, $m_{d,t}\left( x\right) =\mathbb{E}\left[
Y|D=d,T=t,X=x\right] $. In the next section, we differentiate the notation
for the panel data and repeated cross-section case to avoid potential
confusions.} see e.g. \cite{Heckman1997}.

The IPW approach proposed by \cite{Abadie2005} avoids directly modelling the
outcome evolution and exploits that, under Assumptions \ref{ass:sampling}-%
\ref{ass:common.support}, the ATT can be expressed as 
\begin{equation*}
\tau =\frac{1}{\mathbb{E}\left[ D\right] }~\mathbb{E}\left[ \frac{D-p\left(
X\right) }{1-p\left( X\right) }\left( Y_{1}-Y_{0}\right) \right]
\end{equation*}%
when panel data are available, and as 
\begin{equation}
\tau =\frac{1}{\mathbb{E}\left[ D\right] }~\mathbb{E}\left[ \frac{D-p\left(
X\right) }{1-p\left( X\right) }\frac{T-\lambda }{\lambda \left( 1-\lambda
\right) }Y\right]  \label{abadie.rc}
\end{equation}%
when repeated cross-section data are available, where $p\left( X\right)
\equiv \mathbb{P}\left( D=1|X\right) $ is the true, unknown propensity
score. Abadie's identification results suggest simple two-step estimators
for the ATT that do not involve outcome regressions. For instance, when
panel data are available, \cite{Abadie2005} proposes the following \cite%
{Horvitz1952} type IPW estimator, 
\begin{equation}
\widehat{\tau }^{ipw,p}=\frac{1}{\mathbb{E}_{n}\left[ D\right] }~\mathbb{E}%
_{n}\left[ \frac{D-\widehat{\pi }\left( X\right) }{1-\widehat{\pi }(X)}%
\left( Y_{1}-Y_{0}\right) \right] ,  \label{att.ipw}
\end{equation}%
where $\widehat{\pi }\left( x\right) $ is an estimator of the true, unknown $%
p\left( x\right) $, and for a generic random variable $Z$, $\mathbb{E}_{n}%
\left[ Z\right] =n^{-1}\sum_{i=1}^{n}Z_{i};$ the estimator for the repeated
cross-section case is formed using the analogous procedure.

It is important to emphasize that the reliability of ATT estimators based on
the OR and the IPW approaches depends on different, non-nested conditions.
For the OR approach, the consistency of the ATT estimator (\ref{att.reg1})
relies on the estimators of $m_{d,t}\left( \cdot \right) $, $\widehat{\mu }%
_{d,t}\left( \cdot \right) $, being correctly specified, whereas the IPW
estimator (\ref{att.ipw}) relies on the propensity score estimator $\widehat{%
\pi }(\cdot )$ of $p\left( \cdot \right) $ being correctly specified. As so,
in practice, it may be hard to \textquotedblleft rank\textquotedblright\
these two approaches in terms of their robustness to model misspecification.

\begin{remark}
\label{rem:twfe}It is common to see practitioners adopting the two-way fixed
effects linear regression model 
\begin{equation}
Y_{it}=\alpha _{1}+\alpha _{2}~T_{i}+\alpha _{3}~D_{i}+\tau ^{fe}~\left(
T_{i}\cdot D_{i}\right) +\theta ^{\prime }X_{i}+\epsilon _{it},  \label{fe}
\end{equation}%
and interpreting estimates of $\tau ^{fe}$ as estimates of the ATT, see
e.g.~chapter 5.2 in \cite{Angrist2009}. Although (\ref{fe}) may be perceived
as a \textquotedblleft natural\textquotedblright\ specification, it
implicitly imposes additional restrictions on the data generating process
beyond Assumptions \ref{ass:sampling}-\ref{ass:common.support}. More
specifically, (\ref{fe}) implicitly imposes that $\left( i\right) \mathbb{E}%
\left[ Y_{1}\left( 1\right) -Y_{1}\left( 0\right) |X,D=1\right] =\tau
^{fe}~a.s.$, i.e., it assumes homogeneous (in $X$)\ treatment effects, and $%
\left( ii\right) $ for $d=0,1$, $\mathbb{E}\left[ Y_{1}-Y_{0}|X,D=d\right] =%
\mathbb{E}\left[ Y_{1}-Y_{0}|D=d\right] ~a.s.$, i.e., it rules out $X$%
-specific trends in both treated and comparison groups.\footnote{%
Note that under Assumptions \ref{ass:sampling}-\ref{ass:common.support}, (%
\ref{fe}) suggests that, with probability one, $\mathbb{E}\left[ Y_{1}\left(
1\right) |X,D=1\right] =\alpha _{1}+\alpha _{2}+\alpha _{3}+\tau +\theta
^{\prime }X~,$ and $\mathbb{E}\left[ Y_{1}\left( 0\right) |X,D=1\right] =%
\mathbb{E}[Y_{0}|D=1,X]+\left( \mathbb{E}[Y_{1}|D=0,X]-\mathbb{E}%
[Y_{0}|D=0,X]\right) =\alpha _{1}+\alpha _{2}+\alpha _{3}+\theta ^{\prime
}X~ $. Point $\left( i\right) $ now follows directly. Point $\left(
ii\right) $ follows from analogous arguments.} When these additional
restrictions are not satisfied, the estimand $\tau ^{fe}$ is, in general,
different from the ATT, and policy evaluation based on it may be misleading.
We further illustrate this point using Monte Carlo simulations in Section %
\ref{sec:MC}; see also \cite{Sloczynski2017} for related results.
\end{remark}

\subsection{Doubly robust difference-in-differences estimands\label{DR}}

In this section, we argue that instead of choosing between the OR and the
IPW approaches, one can combine them to form doubly robust (DR)
moments/estimands for the ATT. Here, double robustness means that the
resulting estimand identifies the ATT even if either (but not both) the
propensity score model or the outcome regression models are misspecified. As
so, the DR DID estimand for the ATT shares the strengths of each individual
DID method and, at the same time, avoids some of their weaknesses.

Before describing how we exactly combine the OR and the IPW approaches to
form our DR DID estimand, we need to introduce some additional notation. Let 
$\pi \left( X\right) $ be an arbitrary model for the true, unknown
propensity score. When panel data are available, let $\Delta Y=Y_{1}-Y_{0}$
and define $\mu _{d,\Delta }^{p}\left( X\right) \equiv \mu _{d,1}^{p}\left(
X\right) -\mu _{d,0}^{p}\left( X\right) $, $\mu _{d,t}^{p}\left( x\right) $
being a model for the true, unknown outcome regression $m_{d,t}^{p}\left(
x\right) \equiv \mathbb{E}[Y_{t}|D=d,X=x]$, $d,t=0,1$. When only repeated
cross-section data are available, let $\mu _{d,t}^{rc}\left( x\right) $ be
an arbitrary model for the true, unknown regression $m_{d,t}^{rc}\left(
x\right) \equiv \mathbb{E}[Y|D=d,T=t,X=x]$, $d,t=0,1$, and for, $d=0,1,$ $%
\mu _{d,Y}^{rc}\left( T,X\right) \equiv T\cdot \mu _{d,1}^{rc}\left(
X\right) +\left( 1-T\right) \cdot \mu _{d,0}^{rc}\left( X\right) $, and $\mu
_{d,\Delta }^{rc}\left( X\right) \equiv \mu _{d,1}^{rc}\left( X\right) -\mu
_{d,0}^{rc}\left( X\right) $.

For the case in which panel data are available, we consider the estimand%
\begin{equation}
\tau ^{dr,p}=\mathbb{E}\left[ \left( w_{1}^{p}\left( D\right)
-w_{0}^{p}\left( D,X;\pi \right) \right) \left( \Delta Y-\mu _{0,\Delta
}^{p}(X)\right) \right] ,  \label{dr.p}
\end{equation}%
where, for a generic $g$,%
\begin{equation}
w_{1}^{p}\left( D\right) =\frac{D}{\mathbb{E}\left[ D\right] },~~~~\text{and}%
~~~w_{0}^{p}\left( D,X;g\right) =\left. \frac{g(X)\left( 1-D\right) }{1-g(X)}%
\right/ \mathbb{E}\left[ \frac{g(X)\left( 1-D\right) }{1-g(X)}\right] .
\label{weights.p}
\end{equation}%
For the repeated cross-section case, we consider two different estimands, 
\begin{equation}
\tau _{1}^{dr,rc}=\mathbb{E}\left[ \left( w_{1}^{rc}\left( D,T\right)
-w_{0}^{rc}\left( D,T,X;\pi \right) \right) \left( Y-\mu
_{0,Y}^{rc}(T,X)\right) \right] ,  \label{dr.rc1}
\end{equation}%
and%
\begin{multline}
\tau _{2}^{dr,rc}=\tau _{1}^{dr,rc}+\left( \mathbb{E}\left[ \left. \mu
_{1,1}^{rc}\left( X\right) -\mu _{0,1}^{rc}\left( X\right) \right\vert D=1%
\right] -\mathbb{E}\left[ \left. \mu _{1,1}^{rc}\left( X\right) -\mu
_{0,1}^{rc}\left( X\right) \right\vert D=1,T=1\right] \right)  \label{dr.rc2}
\\
-\left( \mathbb{E}\left[ \left. \mu _{1,0}^{rc}\left( X\right) -\mu
_{0,0}^{rc}\left( X\right) \right\vert D=1\right] -\mathbb{E}\left[ \left.
\mu _{1,0}^{rc}\left( X\right) -\mu _{0,0}^{rc}\left( X\right) \right\vert
D=1,T=0\right] \right) ,
\end{multline}%
where, for a generic $g$,%
\begin{equation}
w_{1}^{rc}\left( D,T\right) =w_{1,1}^{rc}\left( D,T\right)
-w_{1,0}^{rc}\left( D,T\right) ,\text{ }~~~~\text{and}~~~w_{0}^{rc}\left(
D,T,X;g\right) =w_{0,1}^{rc}\left( D,T,X;g\right) -w_{0,0}^{rc}\left(
D,T,X;g\right) ,  \label{weights.rc}
\end{equation}%
and, for $t=0,1$,%
\begin{eqnarray*}
w_{1,t}^{rc}\left( D,T\right) &=&\frac{D\cdot 1\left\{ T=t\right\} }{\mathbb{%
E}\left[ D\cdot 1\left\{ T=t\right\} \right] }, \\
w_{0,t}^{rc}\left( D,T,X;g\right) &=&\left. \frac{g(X)\left( 1-D\right)
\cdot 1\left\{ T=t\right\} }{1-g(X)}\right/ \mathbb{E}\left[ \frac{%
g(X)\left( 1-D\right) \cdot 1\left\{ T=t\right\} }{1-g(X)}\right] .
\end{eqnarray*}

\begin{theorem}
\label{th:dr} Let Assumptions \ref{ass:sampling}-\ref{ass:common.support}
hold. Then:

$\left( a\right) ~$When panel data are available, $\tau ^{dr,p}=\tau $ if
either (but not necessarily both) $\pi (X)=p\left( X\right) $ $a.s.$ or $\mu
_{\Delta }^{p}\left( X\right) =m_{0,1}^{p}\left( X\right) -m_{0,0}^{p}\left(
X\right) $ $a.s.;$

$\left( b\right) ~$When repeated cross-section data are available, $\tau
_{1}^{dr,rc}=\tau _{2}^{dr,rc}=\tau $ if either (but not necessarily both) $%
\pi (X)=p\left( X\right) $ $a.s.$ or $\mu _{0,\Delta }^{rc}\left( X\right)
=m_{0,1}^{rc}\left( X\right) -m_{0,0}^{rc}\left( X\right) $ $a.s..$
\end{theorem}

Theorem \ref{th:dr} states that provided that at least one of the working
nuisance models is correctly specified, we can recover the ATT with either
panel or repeated cross-section data. Thus, our proposed DR DID estimands
are \textquotedblleft less demanding\textquotedblright\ in terms of the
researchers' ability to correctly specify models for the nuisance functions
than either the OR or the IPW approach.

Given that we consider two different estimands for the case of repeated
cross-section, it is interesting to use Theorem \ref{th:dr} to compare them.
Given that $\tau _{1}^{dr,rc}$ does not rely on OR models for the treated
group but $\tau _{2}^{dr,rc}$ does, one could a priori expect that $\tau
_{1}^{dr,rc}$ would be more robust\ against model misspecification than $%
\tau _{2}^{dr,rc}$. Nonetheless, Theorem \ref{th:dr} states that this is not
the case as they identify the ATT under the same conditions. At this stage,
one may wonder how this is possible. To answer such a query, it suffices to
remember that, under the stationarity condition in Assumption \ref%
{ass:sampling}$\left( b\right) $, for any generic integrable and measurable
function $g$, $\mathbb{E}\left[ \left. g\left( X\right) \right\vert D=1%
\right] =\mathbb{E}\left[ \left. g\left( X\right) \right\vert D=1,T=t\right] 
$, $t=0,1$. Given that this holds for any generic function $g$, it must also
hold for $\mu _{1,t}^{rc}\left( \cdot \right) -\mu _{0,t}^{rc}\left( \cdot
\right) ,$ $t=0,1$, even when $\mu _{d,t}^{rc}\left( \cdot \right) $ are
misspecified models of $m_{d,t}^{rc}\left( \cdot \right) $. Such a result
reveals that modeling the OR for the treat group can be \textquotedblleft
harmless\textquotedblright\ in terms of identification, provided that these
additional models are incorporated into $\tau _{1}^{dr,rc}$ in an
appropriate manner.

\subsection{Semiparametric efficiency bound}

In the previous subsection, we derived DR moment equations for the ATT under
the DID framework and showed that the resulting estimands are more robust
against model misspecifications than DID estimands based on either the OR or
the IPW approach. In this subsection, we shift our attention from
\textquotedblleft robustness\textquotedblright\ to efficiency. More
precisely, we calculate the semiparametric efficiency bound for the ATT
under Assumptions \ref{ass:sampling}-\ref{ass:common.support} when either
panel or repeated cross-section data are available. These results provide
the semiparametric analog of the Cram\'{e}r--Rao lower bound commonly used
in fully parametric procedures. As so, they provide a benchmark that
researchers can use to assess whether any given (regular) semiparametric DID
estimator for the ATT is fully exploiting the empirical content of
Assumptions \ref{ass:sampling}-\ref{ass:common.support}.

Let $m_{0,\Delta }^{p}\left( x\right) \equiv m_{0,1}^{p}\left( x\right)
-m_{0,0}^{p}\left( x\right) $, and, for $d=0,1,$ $m_{d,\Delta }^{rc}\left(
X\right) \equiv m_{d,1}^{rc}\left( X\right) -m_{d,0}^{rc}\left( X\right) $.
Recall that $\lambda \equiv \mathbb{P}\left( T=1\right) $. Next proposition
displays the semiparametric efficiency bound for the ATT when one has access
to panel data and when one has access to repeated cross-section data. To
simplify exposition, we abstract from additional technical discussions
related to the conditions to guarantee quadratic mean differentiability and
their implications for the precise definition of efficient influence
function ; see, e.g., Chapter 3 of \cite{Bickel1998} for additional details.

\begin{proposition}
\label{th:efficiency} Let Assumptions \ref{ass:sampling}-\ref%
{ass:common.support} hold. Then:

$\left( a\right) $ When panel data are available, the efficient influence
function for the $ATT$ is 
\begin{multline}
\eta ^{e,p}\left( Y_{1},Y_{0},D,X\right) =~w_{1}^{p}\left( D\right) \left(
m_{1,\Delta }^{p}\left( X\right) -m_{0,\Delta }^{p}\left( X\right) -\tau
\right)  \label{eq:eff.score.p} \\
+w_{1}^{p}\left( D\right) \left( \Delta Y-m_{1,\Delta }^{p}(X)\right)
-w_{0}^{p}\left( D,X;p\right) \left( \Delta Y-m_{0,\Delta }^{p}(X)\right) ,
\end{multline}%
and the semiparametric efficiency bound for all regular estimators for the
ATT is%
\begin{multline}
\mathbb{E}\left[ \eta ^{e,p}\left( Y_{1},Y_{0},D,X\right) ^{2}\right] =\frac{%
1}{\mathbb{E}\left[ D\right] ^{2}}\mathbb{E}\left[ D\left( m_{1,\Delta
}^{p}\left( X\right) -m_{0,\Delta }^{p}\left( X\right) -\tau \right)
^{2}\right. \\
\left. +D\left( \Delta Y-m_{1,\Delta }^{p}\left( X\right) \right) ^{2}+%
\dfrac{\left( 1-D\right) p\left( X\right) ^{2}}{\left( 1-p\left( X\right)
\right) ^{2}}\left( \Delta Y-m_{0,\Delta }^{p}\left( X\right) \right) ^{2}%
\right] .  \label{eq:eff.var.p}
\end{multline}

$\left( b\right) $ When only repeated cross-section data are available, the
efficient influence function for the $ATT$ is 
\begin{multline}
\eta ^{e,rc}\left( Y,D,T,X\right) =~\frac{D}{\mathbb{E}\left[ D\right] }%
\left( m_{1,\Delta }^{rc}\left( X\right) -m_{0,\Delta }^{rc}\left( X\right)
-\tau \right) \\
+\left( w_{1,1}^{rc}\left( D,T\right) \left( Y-m_{1,1}^{rc}\left( X\right)
\right) -w_{1,0}^{rc}\left( D,T\right) \left( Y-m_{1,0}^{rc}\left( X\right)
\right) \right)  \label{eq:eff.score.rc} \\
-\left( w_{0,1}^{rc}\left( D,T,X;p\right) \left( Y-m_{0,1}^{rc}\left(
X\right) \right) -w_{0,0}^{rc}\left( D,T,X;p\right) \left(
Y-m_{0,0}^{rc}\left( X\right) \right) \right) ,
\end{multline}

and the semiparametric efficiency bound for all regular estimators for the
ATT is%
\begin{multline}
\mathbb{E}\left[ \eta ^{e,rc}\left( Y,D,T,X\right) ^{2}\right] =\frac{1}{%
\mathbb{E}\left[ D\right] ^{2}}\mathbb{E}\left[ D\left( m_{1,\Delta
}^{rc}\left( X\right) -m_{0,\Delta }^{rc}\left( X\right) -\tau \right)
^{2}\right. \\
+\frac{DT}{\mathbb{\lambda }^{2}}\left( Y-m_{1,1}^{rc}\left( X\right)
\right) ^{2}+\frac{D\left( 1-T\right) }{\left( 1-\mathbb{\lambda }\right)
^{2}}\left( Y-m_{1,0}^{rc}\left( X\right) \right) ^{2}  \label{eq:eff.var.rc}
\\
\left. +\frac{\left( 1-D\right) p\left( X\right) ^{2}T}{\left( 1-p\left(
X\right) \right) ^{2}\lambda ^{2}}\left( Y-m_{0,1}^{rc}(X)\right) ^{2}+\frac{%
\left( 1-D\right) p\left( X\right) ^{2}\left( 1-T\right) }{\left( 1-p\left(
X\right) \right) ^{2}\left( 1-\mathbb{\lambda }\right) ^{2}}\left(
Y-m_{0,0}^{rc}\left( X\right) \right) ^{2}\right] .
\end{multline}
\end{proposition}

It is interesting to compare $\eta ^{e,p}\left( D,X\right) $ with $\eta
^{e,rc}\left( D,T,X\right) $. First, note that the first component of their
efficient influence functions are analogous to each other, and depends on
the true, unknown conditional ATT, $m_{1,\Delta }\left( X\right)
-m_{0,\Delta }\left( X\right) $.\footnote{%
To avoid excessive notational burden, we supress the \textquotedblleft $p$%
\textquotedblright\ and \textquotedblleft $rc$\textquotedblright\
superscripts unless their omission leads to confusion.} The second and third
terms in (\ref{eq:eff.score.p}) and (\ref{eq:eff.score.rc}) are more
different from each other. For $\eta ^{e,p}$, the availability of panel data
implies that $Y_{1}$ and $Y_{0}$ are observed for all units, and, therefore,
we can directly reweight $\Delta Y-m_{1,\Delta }\left( X\right) $ and $%
\Delta Y-m_{0,\Delta }\left( X\right) $. In contrast, when only repeated
cross-section data are available, one observes $Y_{t}$ only if $T=t$, $t=0,1$%
, and, therefore, the efficient influence function (\ref{eq:eff.score.rc})
depends on different weights for each pair $\left( D,T\right) $ $\in \left\{
0,1\right\} ^{2}$. In this latter case, we also stress the importance of
imposing the stationarity condition in Assumption \ref{ass:sampling}(b) when
deriving the efficient influence function (\ref{eq:eff.score.rc}) -- failing
to do so will suggest an \textquotedblleft efficiency
bound\textquotedblright\ that is wider than (\ref{eq:eff.var.rc}).

It is also worth mentioning that the efficient influence functions (\ref%
{eq:eff.score.p}) and (\ref{eq:eff.score.rc}) depend on the true, unknown,
outcome regression functions for the treated group, $m_{1,1}\left( \cdot
\right) $ and $m_{1,0}\left( \cdot \right) $, in an asymmetric manner. On
one hand, when panel data are available, by simple manipulation, we can
rewrite $\eta ^{e,p}$ as%
\begin{equation*}
\eta ^{e,p}\left( Y_{1},Y_{0},D,X\right) =\left( w_{1}^{p}\left( D\right)
-w_{0}^{p}\left( D,X;p\right) \right) \left( \Delta Y-m_{0,\Delta }\left(
X\right) \right) -w_{1}^{p}\left( D\right) \cdot \tau ,
\end{equation*}%
emphasizing that the efficient influence function for the ATT when panel
data are available \emph{does not} depend on $m_{1,1}\left( \cdot \right) $
and $m_{1,0}\left( \cdot \right) $. This is in sharp contrast to the case
where only repeated cross-section data are available.

Another interesting question raised by Proposition \ref{th:efficiency} is
whether the semiparametric efficiency bound for the case of repeated
cross-section data is larger than the one for the case of panel data. In
order to answer this question, we consider the case where $T$ is independent
of $\left( Y_{1},Y_{0},D,X\right) $, so that Assumptions \ref{ass:sampling}$%
\left( a\right) $ and \ref{ass:sampling}$\left( b\right) $ are compatible
with each other.\footnote{%
This \textquotedblleft restriction\textquotedblright\ does not affect the
semiparametric efficiency bound for the case where only repeated
cross-section data are available, as it does not impose additional
restrictions on the observed data.}

\begin{corollary}
\label{cor:var.bound} Let Assumptions \ref{ass:sampling}-\ref%
{ass:common.support} hold, and assume that $T$ is independent of $\left(
Y_{1},Y_{0},D,X\right) $. Then,%
\begin{multline*}
\mathbb{E}\left[ \eta ^{e,rc}\left( Y,D,T,X\right) ^{2}\right] -\mathbb{E}%
\left[ \eta ^{e,p}\left( Y_{1},Y_{0},D,X\right) ^{2}\right] \\
=\frac{1}{\mathbb{E}\left[ D\right] ^{2}}\mathbb{E}\left[ D\left( \sqrt{%
\frac{1-\lambda }{\lambda }}\left( Y_{1}-m_{1,1}\left( X\right) \right) +%
\sqrt{\frac{\lambda }{1-\lambda }}\left( Y_{0}-m_{1,0}\left( X\right)
\right) \right) ^{2}\right. \\
+\left. \frac{\left( 1-D\right) p\left( X\right) ^{2}}{\left( 1-p\left(
X\right) \right) ^{2}}\left( \sqrt{\frac{1-\lambda }{\lambda }}\left(
Y_{1}-m_{0,1}\left( X\right) \right) +\sqrt{\frac{\lambda }{1-\lambda }}%
\left( Y_{0}-m_{0,0}\left( X\right) \right) \right) ^{2}\right] \geq 0.
\end{multline*}
\end{corollary}

In other words, under the DID framework it is possible to form more
efficient estimators for the ATT when panel data are available than when
only repeated cross-section data are available. In addition, from Corollary %
\ref{cor:var.bound}, we can also see that the efficiency loss is convex in $%
\lambda \,$, implying that the loss of efficiency is bigger when the pre and
post-treatment sample sizes are more imbalanced. In fact, when 
\begin{multline}
\mathbb{E}\left[ D\left( Y_{0}-m_{1,0}\left( X\right) \right) ^{2}+\frac{%
\left( 1-D\right) p\left( X\right) ^{2}}{\left( 1-p\left( X\right) \right)
^{2}}\left( Y_{0}-m_{0,0}\left( X\right) \right) ^{2}\right] =
\label{eq:mom.lambda} \\
\mathbb{E}\left[ D\left( Y_{1}-m_{1,1}\left( X\right) \right) ^{2}+\frac{%
\left( 1-D\right) p\left( X\right) ^{2}}{\left( 1-p\left( X\right) \right)
^{2}}\left( Y_{1}-m_{0,1}\left( X\right) \right) ^{2}\right] ,
\end{multline}%
we can show that $\lambda =0.5$ is optimal . However, when (\ref%
{eq:mom.lambda}) does not hold, the optimal $\lambda $ depends on the data
in a more complicated manner, and is given by $\lambda =\left. \tilde{\sigma}%
_{1}\right/ \left( \tilde{\sigma}_{0}+\tilde{\sigma}_{1}\right) $, where,
for $t=0,1$ 
\begin{equation*}
\tilde{\sigma}_{t}^{2}=\mathbb{E}\left[ D\left( Y_{t}-m_{1,t}\left( X\right)
\right) ^{2}+\frac{\left( 1-D\right) p\left( X\right) ^{2}}{\left( 1-p\left(
X\right) \right) ^{2}}\left( Y_{t}-m_{0,t}\left( X\right) \right) ^{2}\right]
.
\end{equation*}%
These results suggest that, in principle, one may benefit from
\textquotedblleft oversampling\textquotedblright\ from either the pre or
post-treatment period. However, it is, in general, not feasible to know the
optimal\ $\lambda $ during the design stage, i.e., at the pre-treatment
period, since $\tilde{\sigma}_{1}^{2}$ depends on the outcome data from the
post-treatment period. Thus, if one were to design the DID study with
repeated cross-section units, it seems that setting $\lambda =0.5$ would be
a \textquotedblleft reasonable\textquotedblright\ choice.

\section{Estimation and inference\label{sec:asy.theory}}

In this section, we build on the DR DID estimands in Theorem \ref{th:dr} and
the semiparametric efficiency bounds in Proposition \ref{th:efficiency}, and
discuss estimation and inference procedures for the ATT in DID designs.
Indeed, the moment equations (\ref{dr.p}), (\ref{dr.rc1}), and (\ref{dr.rc2}%
) suggest a simple two-step strategy to estimate the ATT. In the first step,
one estimates the true, unknown $p\left( \cdot \right) $ using $\pi (\cdot
), $ and the true, unknown $m_{d,t}^{p}\left( \cdot \right) $ ($%
m_{d,t}^{rc}\left( \cdot \right) $) using $\mu _{d,t}^{p}\left( \cdot
\right) $($\mu _{d,t}^{rc}\left( \cdot \right) $), $d,t=0,1$, when panel
data (repeated cross-section data) are available. In the second step, one
plugs the fitted values of the estimated propensity score and regression
models into the sample analogue of $\tau ^{dr,p},$ $\tau _{1}^{dr,rc}$, or $%
\tau _{2}^{dr,rc}$.

Although, in principle, one can use semi/non-parametric estimators for both
the outcome regressions and the propensity score, see e.g.~\cite{Heckman1997}%
, \cite{Abadie2005}, \cite{Chen2008} and \cite{Rothe2018}, in what follows
,we focus our attention on generic parametric first-step estimators. More
precisely, we assume that $\pi \left( x;\gamma ^{\ast }\right) $ is a
parametric model for $p\left( x\right) ,$ such that $\pi $ is known up to
the finite dimensional pseudo-true parameter $\gamma ^{\ast }$. Analogously,
for $d,t=0,1$, $\mu _{d,t}^{p}\left( x;\beta _{d,t}^{\ast ,p}\right) $ (and $%
\mu _{d,t}^{rc}\left( x;\beta _{d,t}^{\ast ,rc}\right) $) is a parametric
model for $m_{d,t}^{p}\left( x\right) $ ($m_{d,t}^{rc}\left( x\right) $),
such that $\mu _{d,t}^{p}$ ($\mu _{d,t}^{rc}$) is known up to the finite
dimensional pseudo-true parameter $\beta _{d,t}^{\ast ,p}$ ($\beta
_{d,t}^{\ast ,rc}$). This is perhaps the most popular approach adopted by
practitioners, particularly when the available sample size is moderate
and/or the dimension of available covariates is high or even moderate, as
the \textquotedblleft curse of dimensionality\textquotedblright\ usually
prevents one to adopt fully nonparametric procedures.\footnote{%
Let $g\left( x\right) $ be a generic notation for $p\left( x\right) $, $%
m_{d,t}^{l}\left( X\right) ,$ $m_{d,t}^{l}\left( X\right) $, $d,t=0,1$, $%
l=p,rc.$ From \cite{Newey1994}, \cite{Chen2003}, \cite{Ai2003, Ai2007,
Ai2012}, and \cite{Chen2008}, one can see that the use of nonparametric
first-step estimators $\widehat{g}\left( x\right) $ of $g\left( x\right) $
is warranted provided that $\left\Vert \widehat{g}\left( x\right) -g\left(
x\right) \right\Vert _{\mathcal{H}}=o_{p}\left( n^{-1/4}\right) $ for a
pseudo-metric $\left\Vert \cdot \right\Vert _{\mathcal{H}}$, $\mathcal{H}$
being a vector space of functions. However, when the dimension of $X$ is
moderate or large, as is usually the case in many empirical applications,
conditons ensuring that $\left\Vert \widehat{g}\left( x\right) -g\left(
x\right) \right\Vert _{\mathcal{H}}=o_{p}\left( n^{-1/4}\right) $ can be
rather stringent because of the \textquotedblleft curse of
dimensionality\textquotedblright .}

In the case when panel data are available, our proposed DR DID estimator for
the ATT is based on (\ref{dr.p}) and is given by 
\begin{equation}
\widehat{\tau }^{dr,p}=\mathbb{E}_{n}\left[ \left( \widehat{w}_{1}^{p}\left(
D\right) -\widehat{w}_{0}^{p}\left( D,X;\widehat{\gamma }\right) \right)
\left( \Delta Y-\mu _{0,\Delta }^{p}\left( X;\widehat{\beta }_{0,0}^{p},%
\widehat{\beta }_{0,1}^{p}\right) \right) \right] ,  \label{dr.p.estimator}
\end{equation}%
where 
\begin{equation}
\widehat{w}_{1}^{p}\left( D\right) =\frac{D}{\mathbb{E}_{n}\left[ D\right] },%
\text{ ~~~~~and~~~~~ }\widehat{w}_{0}^{p}\left( D,X;\gamma \right) =\left. 
\frac{\pi (X;\gamma )\left( 1-D\right) }{1-\pi (X;\gamma )}\right/ \mathbb{E}%
_{n}\left[ \frac{\pi (X;\gamma )\left( 1-D\right) }{1-\pi (X;\gamma ))}%
\right] ,  \label{w.hat}
\end{equation}%
$\widehat{\gamma }$ is an estimator for the pseudo-true $\gamma ^{\ast }$, $%
\widehat{\beta }_{0,t}^{p}$ is an estimator for pseudo-true $\beta
_{0,t}^{\ast ,p}$, $t=0,1$, and for a generic $\beta _{0}$ and $\beta _{1}$, 
$\mu _{0,\Delta }^{p}(\cdot ;\beta _{0},\beta _{1})=\mu _{0,1}^{p}\left(
\cdot ;\beta _{1}\right) -\mu _{0,0}^{p}\left( \cdot ;\beta _{0}\right) $.

When only repeated cross-section data are available, we propose two
different DR DID estimators for the ATT. The first one, which is based on (%
\ref{dr.rc1}) and can be interpreted as the analogue of $\widehat{\tau }%
^{dr,p}$, is given by 
\begin{equation}
\widehat{\tau }_{1}^{dr,rc}=\mathbb{E}_{n}\left[ \left( \widehat{w}%
_{1}^{rc}\left( D,T\right) -\widehat{w}_{0}^{rc}\left( D,T,X;\widehat{\gamma 
}\right) \right) \left( Y-\mu _{0,Y}^{rc}\left( T,X;\widehat{\beta }%
_{0,0}^{rc},\widehat{\beta }_{0,1}^{rc}\right) \right) \right] ,
\label{dr.rc1.estimator}
\end{equation}%
where $\mu _{0,Y}^{rc}\left( T,\cdot ;\beta _{0,0}^{rc},\beta
_{0,1}^{rc}\right) =T\cdot \mu _{0,1}^{rc}\left( \cdot ;\beta
_{0,1}^{rc}\right) +\left( 1-T\right) \cdot \mu _{0,0}^{rc}\left( \cdot
;\beta _{0,0}^{rc}\right) $, $\widehat{\beta }_{d,t}^{rc}$ is an estimator
for the pseudo-true $\beta _{d,t}^{\ast ,rc}$, $d,t=0,1$, and the weights $%
\widehat{w}_{1}^{rc}\left( D,T\right) $ and $\widehat{w}_{0}^{rc}\left(
D,T,X;\widehat{\gamma }\right) $ are, respectively, defined as the sample
analogues of $w_{1}^{rc}\left( D,T\right) $ and $w_{0}^{rc}\left(
D,T,X;g\right) $ defined in (\ref{weights.rc}), but with $\pi \left( x;%
\widehat{\gamma }\right) $ playing the role of $g$.

The second DR DID estimator for the case of repeated cross-section builds on
(\ref{dr.rc2}) and is given by%
\begin{multline}
\widehat{\tau }_{2}^{dr,rc}=\widehat{\tau }_{1}^{dr,rc}+\left( \mathbb{E}_{n}%
\left[ \left( \frac{D}{\mathbb{E}_{n}\left[ D\right] }-\widehat{w}%
_{1,1}^{rc}\left( D,T\right) \right) \left( \mu _{1,1}^{rc}\left( X;\widehat{%
\beta }_{1,1}^{rc}\right) -\mu _{0,1}^{rc}\left( X;\widehat{\beta }%
_{0,1}^{rc}\right) \right) \right] \right)  \label{dr.rc2.estimator} \\
-\left( \mathbb{E}_{n}\left[ \left( \frac{D}{\mathbb{E}_{n}\left[ D\right] }-%
\widehat{w}_{1,0}^{rc}\left( D,T\right) \right) \left( \mu _{1,0}^{rc}\left(
X;\widehat{\beta }_{1,0}^{rc}\right) -\mu _{0,0}^{rc}\left( X;\widehat{\beta 
}_{0,0}^{rc}\right) \right) \right] \right) ,
\end{multline}%
where $\mu _{d,\Delta }^{rc}\left( \cdot ;\beta _{d,1}^{rc},\beta
_{d,0}^{rc}\right) =\mu _{d,1}^{rc}\left( \cdot ;\beta _{d,1}^{rc}\right)
-\mu _{d,0}^{rc}\left( \cdot ;\beta _{d,0}^{rc}\right) $, and the weights $%
\widehat{w}_{1,t}^{rc}\left( D,T\right) $ and $\widehat{w}_{0,t}^{rc}\left(
D,T,X;\widehat{\gamma }\right) $ are, respectively, defined as the sample
analogues of $w_{1,t}^{rc}\left( D,T\right) $ and $w_{0,t}^{rc}\left(
D,T,X;g\right) $, $t=0,1$, defined below (\ref{weights.rc}), but with $\pi
\left( x;\widehat{\gamma }\right) $ playing the role of $g$.

As we show in the Appendix \ref{App:conditions}, it is relatively
straightforward to derive the asymptotic properties of $\widehat{\tau }%
^{dr,p}$, $\widehat{\tau }_{1}^{dr,rc}$ and $\widehat{\tau }_{2}^{dr,rc}$
using generic first-step estimators that satisfy some relatively weak,
high-level conditions; see Theorems \ref{th:att.panel} and \ref{th:att.rc}
in Appendix \ref{App:conditions}. Indeed, Theorem \ref{th:att.panel}
indicates that $\widehat{\tau }^{dr,p}$ is doubly robust, and also locally
semiparametrically efficient, i.e., its asymptotic variance achieves the
semiparametric efficiency bound when the working models for the nuisance
functions are correctly specified. Theorem \ref{th:att.rc} also indicates
that both $\widehat{\tau }_{1}^{dr,rc}$ and $\widehat{\tau }_{2}^{dr,rc}$
are doubly robust when repeated cross-section data are available. However,
Theorem \ref{th:att.rc} also highlights that $\widehat{\tau }_{2}^{dr,rc}$
is locally semiparametrically efficient, whereas $\widehat{\tau }%
_{1}^{dr,rc} $ is not. In other words, when repeated cross-section data are
available, $\widehat{\tau }_{2}^{dr,rc}$ tends to have more attractive
properties than $\widehat{\tau }_{1}^{dr,rc}$, regardless of the first-step
estimators used.

Although the results in Theorems \ref{th:att.panel} and \ref{th:att.rc}
accommodate a variety of different first-step estimators, in practice, one
still needs to choose a particular estimation procedure to be implemented.
In what follows, we attempt to provide some guidance on the choice of
first-step estimators with the goal of further improving the (generic) DR
DID estimators. We are particularly interested in forming DR DID estimators
that are not only doubly robust in terms of consistency---like described
above---but also doubly robust for inference, i.e., their asymptotic linear
representation is also doubly robust. The attractiveness of forming
estimators that are DR for inference is that there is no estimation effect
from first-step estimators, which, in turn, implies that the asymptotic
variance of the results DR DID estimator for the ATT is invariant to which
working models for the nuisance functions are correctly specified. In
practice, this usually translates to simpler and more stable inference
procedures.

To derive these improved DR DID estimators, we focus on the case where a
researcher is comfortable with linear regression working models for the
outcome of interest, a logistic working model for the propensity score, and
with covariates $X$ entering all the nuisance models in a symmetric manner.
Although these modelling conditions are more stringent than those allowed by
our generic DR DID estimators discussed in Appendix \ref{App:conditions},
they are much weaker than those implicitly imposed in the TWFE specification
(\ref{fe}), and can be seen as the default choice in many applications.
Hence, these extra assumptions can be seen as a reasonable compromise to get
further improved DR DID estimators that are also computationally tractable
and easy to implement in practice.

\subsection{Improved DR DID estimators when panel data are available\label%
{sec:firststep}}

As discussed above, we consider the following working models for the
nuisance functions: 
\begin{equation}
\pi \left( X,\gamma \right) =\Lambda \left( X^{\prime }\gamma \right) \equiv 
\frac{\exp \left( X^{\prime }\gamma \right) }{1+\exp \left( X^{\prime
}\gamma \right) },\text{ and }\mu _{0,\Delta }^{p}\left( X;\beta
_{0,1}^{p},\beta _{0,1}^{p}\right) =\mu _{0,\Delta }^{lin,p}\left( X;\beta
_{0,\Delta }^{p}\right) \equiv X^{\prime }\beta _{0,\Delta }^{p}.
\label{eq:improved.models}
\end{equation}%
Our proposed improved DR DID estimator is given by the three-step estimator 
\begin{equation*}
\widehat{\tau }_{imp}^{dr,p}=\mathbb{E}_{n}\left[ \left( \widehat{w}%
_{1}^{p}\left( D\right) -\widehat{w}_{0}^{p}\left( D,X;\widehat{\gamma }%
^{ipt}\right) \right) \left( \Delta Y-\mu _{0,\Delta }^{lin,p}\left( X;%
\widehat{\beta }_{0,\Delta }^{wls,p}\right) \right) \right] ,
\end{equation*}%
where the first two-steps consist of computing 
\begin{eqnarray*}
\widehat{\gamma }^{ipt} &=&\arg \max_{\gamma \in \Gamma }\mathbb{E}_{n}\left[
DX^{\prime }\gamma -\left( 1-D\right) \exp \left( X^{\prime }\gamma \right) %
\right] , \\
\widehat{\beta }_{0,\Delta }^{wls,p} &=&\arg \min_{b\in \Theta }\mathbb{E}%
_{n}\left[ \left. \frac{\Lambda \left( X^{\prime }\widehat{\gamma }%
^{ipt}\right) }{1-\Lambda \left( X^{\prime }\widehat{\gamma }^{ipt}\right) }%
\left( \Delta Y-X^{\prime }b\right) ^{2}\right\vert D=0\right] ,
\end{eqnarray*}%
while in the third and last step, one plugs the fitted values of the working
models (\ref{eq:improved.models}) into the sample analogue of $\tau ^{dr,p}$%
. {Here, note that }$\widehat{\gamma }^{ipt}$ is the inverse probability
tilting estimator proposed by \cite{Graham2012} in a different context,
while $\widehat{\beta }_{0,\Delta }^{wls,p}$ is simply the weighted least
squares estimator for $\beta _{0,\Delta }^{\ast ,p}$.

At this point, one may wonder why we use the estimators $\widehat{\gamma }%
^{ipt}$ and $\widehat{\beta }_{0,\Delta }^{wls,p}$ instead of other
available alternatives. To answer such a query, recall that the main goal
here is to propose DID estimators for the ATT that are not only DR
consistent but also DR for inference, i.e., the exact form of their
asymptotic variance does not depend on which working models for the nuisance
functions are correctly specified. As it turns out, the key to obtain DID
estimators for the ATT that are also DR\ for inference is to choose
first-step estimators for the nuisance parameters, say $\widehat{\gamma }$
and $\widehat{\beta }$, such that the limiting distribution of the resulting
DR DID estimator $\widehat{\tau }^{dr,p}$ is equivalent to that of the
infeasible DR DID estimator that uses the pseudo-true values of $\widehat{%
\gamma }$ and $\widehat{\beta }$, say $\gamma ^{\ast }$ and $\beta ^{\ast }$%
. In a more precise manner, in order to get DID estimators that are DR for
inference, we need to guarantee that there will be no estimation effect from
the first stage.

In Appendix \ref{App:conditions}, we show that the estimation effect
associated with using generic first-step estimators $\widehat{\gamma }$ and $%
\widehat{\beta }$ is given by $\eta _{est}^{p}\left( W;\gamma ^{\ast },\beta
^{\ast }\right) $ as defined in (\ref{eq:est.p}). By paying closer attention
to the exact form of $\eta _{est}^{p}\left( W;\gamma ^{\ast },\beta ^{\ast
}\right) $, one can see that if 
\begin{eqnarray}
\mathbb{E}\left[ \left( w_{1}^{p}-w_{0}^{p}\left( \gamma ^{\ast }\right)
\right) \cdot \dot{\mu}_{0,\Delta }^{p}\left( \beta ^{\ast }\right) \right] 
&=&0,  \notag \\
\mathbb{E}\left[ \dfrac{\left( 1-D\right) }{\left( 1-\pi (X;\gamma ^{\ast
})\right) ^{2}}\left( \Delta Y-\mu _{0,\Delta }^{p}\left( \beta ^{\ast
}\right) \right) \cdot \dot{\pi}(\gamma ^{\ast })\right]  &=&0,
\label{eq:moments} \\
\mathbb{E}\left[ w_{0}^{p}\left( \gamma ^{\ast }\right) \cdot \left( \Delta
Y-\mu _{0,\Delta }^{p}\left( \beta ^{\ast }\right) \right) \right]  &=&0, 
\notag
\end{eqnarray}%
then there will be no estimation effect from the first stage. As the first
component of $X$ is assumed to be constant and we adopt the working models (%
\ref{eq:improved.models}), it follows that (\ref{eq:moments}) reduces to%
\begin{eqnarray*}
\mathbb{E}\left[ \left( \frac{D}{\mathbb{E}\left[ D\right] }-\frac{\exp
\left( X^{\prime }\gamma ^{\ast }\right) \left( 1-D\right) }{\mathbb{E}\left[
\exp \left( X^{\prime }\gamma ^{\ast }\right) \left( 1-D\right) \right] }%
\right) X\right]  &=&0, \\
\mathbb{E}\left[ \left. \exp \left( X^{\prime }\gamma ^{\ast }\right) \left(
\Delta Y-\mu _{0,\Delta }^{lin,p}\left( X;\beta _{0,\Delta }^{\ast }\right)
\right) X\right\vert D=0\right]  &=&0.
\end{eqnarray*}%
However, as $n\rightarrow \infty $, these two vectors of moment conditions
follow from the first-order conditions of the optimization problems
associated with $\widehat{\gamma }^{ipt}$ and $\widehat{\beta }_{0,\Delta
}^{wls,p}$, respectively, even when these working models are misspecified.
Hence, by using $\widehat{\gamma }^{ipt}$ and $\widehat{\beta }_{0,\Delta
}^{wls,p}$, we guarantee that $\widehat{\tau }_{imp}^{dr,p}$ is doubly
robust for inference as there is no estimation effect from replacing the
pseudo-true parameters $\gamma ^{\ast ,ipt}$ and $\beta _{0,\Delta }^{\ast
,wls,p}$ with their estimators $\widehat{\gamma }^{ipt}$ and $\widehat{\beta 
}_{0,\Delta }^{wls,p}$, respectively.

The next theorem formalizes this discussion. Define%
\begin{eqnarray}
\widehat{\tau }_{imp}^{dr,p} &=&\mathbb{E}_{n}\left[ \left( w_{1}^{p}\left(
D\right) -w_{0}^{p}\left( D,X;\widehat{\gamma }^{ipt}\right) \right) \left(
\Delta Y-\mu _{0,\Delta }^{lin,p}\left( X;\widehat{\beta }_{0,\Delta
}^{wls,p}\right) \right) \right] ,  \label{dr.p.imp.estimator} \\
\tau _{imp}^{dr,p} &=&\mathbb{E}\left[ \left( w_{1}^{p}\left( D\right)
-w_{0}^{p}\left( D,X;\gamma ^{\ast ,ipt}\right) \right) \left( \Delta Y-\mu
_{0,\Delta }^{lin,p}\left( X;\beta _{0,\Delta }^{\ast ,wls,p}\right) \right) %
\right] ,  \notag
\end{eqnarray}%
and let%
\begin{equation*}
\eta _{imp}^{dr,p}\left( W;\gamma ^{\ast ,ipt},\beta _{0,\Delta }^{\ast
,wls,p},\tau _{imp}^{dr,p}\right) =\left( w_{1}^{p}\left( D\right)
-w_{0}^{p}\left( D,X;\gamma ^{\ast ,ipt}\right) \right) \left( \Delta Y-\mu
_{0,\Delta }^{lin,p}\left( X;\beta _{0,\Delta }^{\ast ,wls,p}\right) \right)
-w_{1}^{p}\left( D\right) \cdot \tau _{imp}^{dr,p}.
\end{equation*}

\begin{theorem}
\label{prop:att.br.panel} Suppose Assumptions \ref{ass:sampling}-\ref%
{ass:common.support} and Assumptions \ref{ass:parametric}-\ref%
{ass:integrabilitity} stated in Appendix \ref{App:conditions} hold, and that
the working nuisance models (\ref{eq:improved.models}) are adopted. Then,

$\left( a\right) $ If either $\Lambda \left( X^{\prime }\gamma ^{\ast
,ipt}\right) =p\left( X\right) ~a.s$ or $X^{\prime }\beta _{0,\Delta }^{\ast
,wls,p}=m_{0,\Delta }^{p}\left( X\right) ~a.s.$, then, as $n\rightarrow
\infty $, 
\begin{equation*}
\widehat{\tau }_{imp}^{dr,p}\overset{p}{\rightarrow }\tau ,
\end{equation*}%
and%
\begin{eqnarray*}
\sqrt{n}(\widehat{\tau }_{imp}^{dr,p}-\tau _{imp}^{dr,p}) &=&\frac{1}{\sqrt{n%
}}\sum_{i=1}^{n}\eta _{imp}^{dr,p}\left( W;\gamma ^{\ast ,ipt},\beta
_{0,\Delta }^{\ast ,wls,p},\tau _{imp}^{dr,p}\right) +o_{p}(1) \\
&&\overset{d}{\rightarrow }N\left( 0,V_{imp}^{p}\right) ,
\end{eqnarray*}%
where $V_{imp}^{p}=\mathbb{E}\left[ \eta _{imp}^{dr,p}\left( W;\gamma ^{\ast
,ipt},\beta _{0,\Delta }^{\ast ,wls,p},\tau _{imp}^{dr,p}\right) ^{2}\right] 
$.

$\left( b\right) $If both $\Lambda \left( X^{\prime }\gamma ^{\ast
,ipt}\right) =p\left( X\right) ~a.s$ and $X^{\prime }\beta _{0,\Delta
}^{\ast ,wls,p}=m_{0,\Delta }^{p}\left( X\right) ~a.s.$, then $\eta
_{imp}^{dr,p}\left( W;\gamma ^{\ast ,ipt},\beta _{0,\Delta }^{\ast
,wls,p},\tau _{imp}^{dr,p}\right) =\eta ^{e,p}\left( Y_{1},Y_{0},D,X\right)
~a.s.$ and $V_{imp}^{p}$ is equal to the semiparametrically efficiency bound
(\ref{eq:eff.var.p}).
\end{theorem}

Part $\left( a\right) $ of Theorem \ref{prop:att.br.panel} generalizes the
cross-section results of \cite{Vermeulen2015} to the DID framework. It
states that the proposed DR DID estimators for the ATT, $\widehat{\tau }%
_{imp}^{dr,p}$, is doubly robust, $\sqrt{n}$-consistent and asymptotically
normal. It also states that the exact form of $V_{imp}^{p}$ does not depend
on which working models are correctly specified, implying that $\widehat{%
\tau }_{imp}^{dr,p}$ is doubly robust not only in terms of consistency but
also terms of inference. An important consequence of this DR-for-inference
property is that it allows one to treat the summands of $\widehat{\tau }%
_{imp}^{dr,p}$ as if they were independent and identically distributed, and,
therefore, estimate $V_{imp}^{p}$ by 
\begin{equation*}
\widehat{V}_{imp}^{p}=\mathbb{E}_{n}\left[ \eta _{imp}^{dr,p}\left( W;%
\widehat{\gamma }^{ipt},\widehat{\beta }_{0,\Delta }^{wls,p},\widehat{\tau }%
_{imp}^{dr,p}\right) ^{2}\right] .
\end{equation*}

Part $\left( b\right) $ of Theorem \ref{prop:att.br.panel} indicates that $%
\widehat{\tau }_{imp}^{dr,p}$ is semiparametrically efficient when the
working model for the propensity score, and the working models for the
outcome regression for the comparison units are correctly specified.

\begin{remark}
\label{rem:GMM}From the discussion above, it may be natural to directly use
the moment conditions (\ref{eq:moments}) to form (generic) nonlinear
generalized method of moment (GMM) estimators for $\gamma $ and $\beta $.
However, it is important to emphasize that to justify the use of such
estimation procedure, one must at least establish the local identification
of the pseudo-true parameters, which, in turn, requires the matrix of
derivatives of (\ref{eq:moments}) having full column rank. Importantly, such
a condition may not hold for some working models. This is particularly the
case when one adopts the working models (\ref{eq:improved.models}) and both
specifications are correctly specified. Thus, care must be taken when one
attempts to use alternative, more general estimation techniques to
generalize the DR inference results discussed above.
\end{remark}

\begin{remark}
As discussed in Appendix A of \cite{Graham2012}, it is possible to use
alternative specifications for the propensity score, e.g., a probit working
model. However, when one deviates from the logit specification, the
optimization algorithm involved to estimate the nuisance parameters $\gamma $
tends to be more computationally demanding, as it involves numerical
integration. As discussed above, $\widehat{\gamma }^{ipt}$ clearly avoids
such complications.
\end{remark}

\subsection{Improved DR DID estimators when repeated cross-section data are
available\label{App:improved.rc}}

In this section, we turn our attention to our proposed improved DR DID
estimators for the ATT when only repeated cross-section data are available.
Similar to the panel data case, we consider the case where a researcher is
comfortable with the following specifications, 
\begin{equation}
\pi \left( X,\gamma \right) =\Lambda \left( X^{\prime }\gamma \right) \equiv 
\frac{\exp \left( X^{\prime }\gamma \right) }{1+\exp \left( X^{\prime
}\gamma \right) },\text{ and }\mu _{d,t}^{rc}\left( X;\beta
_{d,t}^{rc}\right) =\mu _{d,t}^{lin,rc}\left( X;\beta _{d,t}^{rc}\right)
\equiv X^{\prime }\beta _{d,t}^{rc}.  \label{eq:improved.models.rc}
\end{equation}

We consider two improved DR DID estimators based on (\ref{dr.rc1}) and (\ref%
{dr.rc2}), namely 
\begin{equation}
\widehat{\tau }_{1,imp}^{dr,rc}=\mathbb{E}_{n}\left[ \left( \widehat{w}%
_{1}^{rc}\left( D,T\right) -\widehat{w}_{0}^{rc}\left( D,T,X;\widehat{\gamma 
}^{ipt}\right) \right) \left( Y-\mu _{0,Y}^{lin,rc}\left( X;\widehat{\beta }%
_{0,1}^{wls,rc},\widehat{\beta }_{0,0}^{wls,rc}\right) \right) \right] ,
\label{dr.rc.imp1.estimator}
\end{equation}%
and%
\begin{multline}
\widehat{\tau }_{2,imp}^{dr,rc}=\widehat{\tau }_{1,imp}^{dr,rc}+\left( 
\mathbb{E}_{n}\left[ \left( \frac{D}{\mathbb{E}_{n}\left[ D\right] }-%
\widehat{w}_{1,1}^{rc}\left( D,T\right) \right) \left( \mu _{1,1}^{rc}\left(
X;\widehat{\beta }_{1,1}^{ols,rc}\right) -\mu _{0,1}^{rc}\left( X;\widehat{%
\beta }_{0,1}^{wls,rc}\right) \right) \right] \right)
\label{dr.rc.imp2.estimator} \\
-\left( \mathbb{E}_{n}\left[ \left( \frac{D}{\mathbb{E}_{n}\left[ D\right] }-%
\widehat{w}_{1,0}^{rc}\left( D,T\right) \right) \left( \mu _{1,0}^{rc}\left(
X;\widehat{\beta }_{1,0}^{ols,rc}\right) -\mu _{0,0}^{rc}\left( X;\widehat{%
\beta }_{0,0}^{wls,rc}\right) \right) \right] \right) ,
\end{multline}%
where 
\begin{eqnarray*}
\widehat{\gamma }^{ipt} &=&\arg \max_{\gamma \in \Gamma }\mathbb{E}_{n}\left[
DX^{\prime }\gamma -\left( 1-D\right) \exp \left( X^{\prime }\gamma \right) %
\right] , \\
\widehat{\beta }_{0,t}^{wls,rc} &=&\arg \min_{b\in \Theta }\mathbb{E}_{n}%
\left[ \left. \frac{\Lambda \left( X^{\prime }\widehat{\gamma }^{ipt}\right) 
}{1-\Lambda \left( X^{\prime }\widehat{\gamma }^{ipt}\right) }\left(
Y-X^{\prime }b\right) ^{2}\right\vert D=0,T=t\right] , \\
\widehat{\beta }_{1,t}^{ols,rc} &=&\arg \min_{b\in \Theta }\mathbb{E}_{n}%
\left[ \left. \left( Y-X^{\prime }b\right) ^{2}\right\vert D=1,T=t\right] .
\end{eqnarray*}%
Here, note that $\widehat{\tau }_{1,imp}^{dr,rc}$ does not rely on OR models
for the treated group while $\widehat{\tau }_{2,imp}^{dr,rc}$ does. In
addition, when one compares $\widehat{\tau }_{1,imp}^{dr,rc}$ and $\widehat{%
\tau }_{2,imp}^{dr,rc}$ with $\widehat{\tau }_{imp}^{dr,p}$, it is evident
that the latter relies on a single OR model since we observe $Y_{1}$ and $%
Y_{0}$ for all units; when only repeated cross-section data are available,
one needs to model the OR in each time period (and each treatment group).
Another interesting feature worth mentioning is that we estimate the OR
parameters for the treated group via ordinary least squares, whereas we
estimate the OR parameters for the control group with weighted least
squares. This follows from the fact that estimating the pseudo-true
parameters $\beta _{1,t}^{\ast ,rc}$, $t=0,1$, does not lead to any
estimation effect, and therefore one can choose her favorite estimation
method. Given this observation and the linear specification in (\ref%
{eq:improved.models.rc}), we find it natural to estimate $\beta _{1,t}^{\ast
,rc}$, $t=0,1$, via OLS as this is the most widespread estimation procedure
adopted by practitioners.

Let 
\begin{equation*}
\tau _{imp}^{dr,rc}=\mathbb{E}\left[ \left( w_{1}^{rc}\left( D,T\right)
-w_{0}^{rc}\left( D,T,X;\gamma ^{\ast ,ipt}\right) \right) \left( Y-\mu
_{0,Y}^{lin,rc}\left( T,X;\beta _{0,1}^{\ast ,wls,rc},\beta _{0,0}^{\ast
,wls,rc}\right) \right) \right]
\end{equation*}%
and for $\beta _{imp}^{\ast ,rc}=\left( \beta _{0,1}^{\ast ,wls,rc},\beta
_{0,0}^{\ast ,wls,rc},\beta _{1,1}^{\ast ,ols,rc},\beta _{1,0}^{\ast
,ols,rc}\right) ,$ define%
\begin{eqnarray*}
\eta _{1,imp}^{dr,rc}\left( W;\gamma ^{\ast ,ipt},\beta _{imp}^{\ast
,rc}\right) &=&\eta _{1}^{rc,1}(W;\beta _{0,1}^{\ast ,wls,rc},\beta
_{0,0}^{\ast ,wls,rc})-\eta _{0}^{rc,1}(W;\gamma ^{\ast ,ipt},\beta
_{0,1}^{\ast ,wls,rc},\beta _{0,0}^{\ast ,wls,rc}), \\
\eta _{2,imp}^{dr,rc}\left( W;\gamma ^{\ast ,ipt},\beta _{imp}^{\ast
,rc}\right) &=&\eta _{1}^{rc,2}(W;\beta _{imp}^{\ast ,rc})-\eta
_{0}^{rc,2}(W;\gamma ^{\ast ,ipt},\beta _{0,1}^{\ast ,wls,rc},\beta
_{0,0}^{\ast ,wls,rc}),
\end{eqnarray*}%
where $\eta _{1}^{rc,1}$, $\eta _{0}^{rc,1}$, $\eta _{1}^{rc,2}$, and $\eta
_{0}^{rc,2}$ are defined as in (\ref{eq:eta1.1.rc})-(\ref{eq:eta.2.0.rc}) in
the Appendix \ref{App:inf.rc}.

Next theorem states that $\widehat{\tau }_{1,imp}^{dr,rc}$ and $\widehat{%
\tau }_{2,imp}^{dr,rc}$ are not only doubly robust consistent but also
doubly robust for inference. Furthermore, it states that $\widehat{\tau }%
_{2,imp}^{dr,rc}$ is locally semiparametrically efficient, whereas $\widehat{%
\tau }_{1,imp}^{dr,rc}$ is not.

\begin{theorem}
\label{prop:att.br.rc}Let $n=n_{1}+n_{0}$, where $n_{1}$ and $n_{0}$ are the
sample sizes of the post-treatment and pre-treatment periods, respectively.
Suppose Assumptions \ref{ass:sampling}-\ref{ass:common.support} and
Assumptions \ref{ass:parametric}-\ref{ass:integrabilitity} stated in
Appendix \ref{App:conditions} hold, that $n_{1}/n\overset{p}{\rightarrow }%
\lambda \in \left( 0,1\right) $ as $n_{0},n_{1}\rightarrow \infty $, and
that the working nuisance models (\ref{eq:improved.models.rc}) are adopted.
Then,

$\left( a\right) $ If either $\Lambda \left( X^{\prime }\gamma ^{\ast
,ipt}\right) =p\left( X\right) ~a.s$ or $X^{\prime }\beta _{0,1}^{\ast
,wls,rc}-X^{\prime }\beta _{0,0}^{\ast ,wls,rc}=m_{0,\Delta }^{rc}\left(
X\right) ~a.s.$, then, for $j=1,2$, as $n\rightarrow \infty $, 
\begin{equation*}
\widehat{\tau }_{j,imp}^{dr,rc}\overset{p}{\rightarrow }\tau ,
\end{equation*}%
and%
\begin{eqnarray*}
\sqrt{n}(\widehat{\tau }_{j,imp}^{dr,rc}-\tau ) &=&\frac{1}{\sqrt{n}}%
\sum_{i=1}^{n}\eta _{j,imp}^{dr,rc}\left( W;\gamma ^{\ast ,ipt},\beta
_{imp}^{\ast ,rc}\right) +o_{p}(1) \\
&&\overset{d}{\rightarrow }N\left( 0,V_{j,imp}^{rc}\right) ,
\end{eqnarray*}%
where $V_{j,imp}^{rc}=\mathbb{E}\left[ \eta _{j,imp}^{dr,rc}\left( W;\gamma
^{\ast ,ipt},\beta _{imp}^{\ast ,rc}\right) ^{2}\right] $.

$\left( b\right) $ Suppose that $\Lambda \left( X^{\prime }\gamma ^{\ast
,ipt}\right) =p\left( X\right) ~a.s$ and, for all $\left( d,t\right) \in
\left\{ 0,1\right\} ^{2}$, $X^{\prime }\beta _{d,t}^{\ast
,wls,rc}=m_{d,t}^{rc}\left( X\right) ~a.s.$. Then, $\eta
_{2,imp}^{dr,rc}\left( W;\gamma ^{\ast ,ipt},\beta _{imp}^{\ast ,rc}\right)
=\eta ^{e,rc}\left( Y,D,T,X\right) $ $a.s.$, and $V_{2,imp}^{rc}$ is equal
to the semiparametrically efficiency bound (\ref{eq:eff.var.rc}). On the
other hand, $V_{1,imp}^{rc}\ $does not attain the semiparametric efficiency
bound.
\end{theorem}

In other words, Theorem \ref{prop:att.br.rc} states that both $\widehat{\tau 
}_{1,imp}^{dr,rc}$ and $\widehat{\tau }_{2,imp}^{dr,rc}$ are doubly robust
for the ATT, $\sqrt{n}$-consistent and asymptotically normal. Similar to the
panel data case, the exact form of the $V_{j,imp}^{rc}$, $j=1,2$, does not
depend on which working models are correctly specified, implying that both $%
\widehat{\tau }_{1,imp}^{dr,rc}$ and $\widehat{\tau }_{2,imp}^{dr,rc}$ are
also doubly robust in terms of inference.

Part $\left( b\right) $ of Theorem \ref{prop:att.br.rc} indicates that $%
\widehat{\tau }_{2,imp}^{dr,rc}$ is semiparametrically efficient when the
working model for the propensity score, and all working models for the
outcome regressions, for both treated and comparison units, are correctly
specified. When compared to Theorem \ref{prop:att.br.panel}$\left( b\right) $%
, it is evident that such a requirement is much stronger than when panel
data are available. Part $\left( b\right) $ of Theorem \ref{prop:att.br.rc}
also indicates that, in general, $\widehat{\tau }_{1,imp}^{dr,rc}$ is not
locally semiparametrically efficient. As so, we argue that, in practice, one
should favor $\widehat{\tau }_{2,imp}^{dr,rc}$ with respect to $\widehat{%
\tau }_{1,imp}^{dr,rc}$, as both estimators are doubly robust in terms of
consistency and inference, but the former is locally semiparametrically
efficiency whereas the latter is not.

We conclude this section by providing a precise characterization of the
efficiency loss associated with using $\widehat{\tau }_{1,imp}^{dr,rc}$
instead of $\widehat{\tau }_{2,imp}^{dr,rc}$ when all working models are
correctly specified. Here, our main goal is to illustrate that by using an
estimator that attempts to mimic the panel data setup and that does not
explicitly exploit the stationarity condition in Assumption \ref%
{ass:sampling}(b), one may incur in substantial efficiency loss. As so, we
argue that, in practice, one should favor estimators based on the DR moment (%
\ref{dr.rc2})---such as $\widehat{\tau }_{2,imp}^{dr,rc}$---with respect to
estimators based on the DR moment (\ref{dr.rc1})---such as $\widehat{\tau }%
_{1,imp}^{dr,rc}$.

\begin{corollary}
\label{cor:var.rc}Suppose the assumptions in Theorem \ref{prop:att.br.rc}
hold. Furthermore, assume that $\Lambda \left( X^{\prime }\gamma ^{\ast
,ipt}\right) =p\left( X\right) ~a.s$ and, for all $\left( d,t\right) \in
\left\{ 0,1\right\} ^{2}$, $X^{\prime }\beta _{d,t}^{\ast
,wls,rc}=m_{d,t}^{rc}\left( X\right) ~a.s.$. Then, 
\begin{equation*}
V_{1,imp}^{rc}-V_{2,imp}^{rc}=\mathbb{E}\left[ D\right] ^{-1}\cdot Var\left[
\left. \sqrt{\frac{1-\lambda }{\lambda }}\left(
m_{1,1}^{rc}(X)-m_{0,1}^{rc}(X)\right) +\sqrt{\frac{\lambda }{1-\lambda }}%
\left( m_{1,0}^{rc}(X)-m_{0,0}^{rc}(X)\right) \right\vert D=1\right] \geq 0.
\end{equation*}
\end{corollary}

\begin{remark}
\label{rem:efficiency}We stress that the result in Corollary \ref{cor:var.rc}
does not depend on the fact that one is using the specifications in (\ref%
{eq:improved.models.rc}). As we show in its proof, such a result remains
true provided that the (generic) first-step estimators for the nuisance
functions are correctly specified. Thus, Corollary \ref{cor:var.rc}
quantifies the loss of efficiency associated with using estimators based on $%
\tau _{1}^{dr,rc}$ as defined in (\ref{dr.rc1})---which includes the
estimator proposed by \cite{Zimmert2019}--- instead of using estimators
based on $\tau _{2}^{dr,rc}$ as defined in (\ref{dr.rc2}). Given that this
loss of efficiency is usually strictly positive, estimators based on $\tau
_{1}^{dr,rc}$ are not, in general, semiparametrically efficient. As we show
in the next section via Monte Carlo simulations, this loss of efficiency can
be large.
\end{remark}

\section{Monte Carlo simulation study\label{sec:MC}}

In this section, we conduct a series of Monte Carlo experiments in order to
study the finite sample properties of our proposed DR DID estimators. When
panel data are available, we compare our proposed DR DID estimators $%
\widehat{\tau }^{dr,p}$ and $\widehat{\tau }_{imp}^{dr,p}$ given in (\ref%
{dr.p.estimator}) and (\ref{dr.p.imp.estimator}), respectively, to the OR
DID estimator (\ref{att.reg1}), the \cite{Horvitz1952} type IPW estimator (%
\ref{att.ipw}), and the TWFE regression model (\ref{fe}). Given that the
weights of the IPW estimator (\ref{att.ipw}) are not normalized to sum up to
one, $\widehat{\tau }^{ipw,p}$ can be unstable particularly when propensity
score estimates are relatively close to one. To assess the role played by
the weights, we also consider the \cite{Hajek1971} type IPW estimator for
the ATT%
\begin{equation}
\widehat{\tau }_{std}^{ipw,p}=\mathbb{E}_{n}\left[ \left( \widehat{w}%
_{1}^{p}\left( D\right) -\widehat{w}_{0}^{p}\left( D,X;\widehat{\gamma }%
\right) \right) \left( Y_{1}-Y_{0}\right) \right] ,  \label{att.ipw2}
\end{equation}%
where the weights $\widehat{w}_{1}^{p}\left( D\right) $ and $\widehat{w}%
_{0}^{p}\left( D,X;\widehat{\gamma }\right) $ are given by (\ref{w.hat}) and
are normalized to sum up to one.

When only repeated cross-section data are available, we compare our proposed
DR DID estimators $\widehat{\tau }_{1}^{dr,rc}$ and $\widehat{\tau }%
_{2}^{dr,rc}$ given in (\ref{dr.rc1.estimator}) and (\ref{dr.rc2.estimator}%
), and their further improved versions $\widehat{\tau }_{1,imp}^{dr,rc}$ and 
$\widehat{\tau }_{2,imp}^{dr,rc}$ given in (\ref{dr.rc.imp1.estimator}) and (%
\ref{dr.rc.imp2.estimator}), to the OR DID estimator (\ref{att.reg1}), the
plug-in IPW estimator based on (\ref{abadie.rc}), and the TWFE regression
model (\ref{fe}). As in the case of panel data, we also consider the \cite%
{Hajek1971} type IPW estimator for the ATT%
\begin{equation}
\widehat{\tau }_{std}^{ipw,rc}=\mathbb{E}_{n}\left[ \left( \widehat{w}%
_{1}^{rc}\left( D,T\right) -\widehat{w}_{0}^{rc}\left( D,T,X;\widehat{\gamma 
}\right) \right) Y\right] ,  \label{att.ipw2.rc}
\end{equation}%
where the weights are the same as those in $\widehat{\tau }_{1}^{dr,rc}$.

In all simulation exercises, we consider a logistic propensity score working
model and a linear regression working model for the outcome evolution. All
observed covariates enter the working models linearly. With the exception of 
$\widehat{\tau }_{imp}^{dr,p}$, $\widehat{\tau }_{j,imp}^{dr,rc}$, $j=1,2$,
where we use the estimation methods proposed in Section \ref{sec:firststep}
and in Section \ref{App:improved.rc}, the OR models are estimated using
ordinary least squares, and the propensity score working model is estimated
using maximum likelihood estimation. When panel data are available, we
consider OR models for $\Delta Y\,$ instead of OR models for $Y_{0}$ and $%
Y_{1}$ separately.

We consider sample size $n$ equal to $1000$. For each design, we conduct $%
10,000$ Monte Carlo simulations. We compare the various DID estimators for
the ATT in terms of average bias, median bias, root mean square error
(RMSE), empirical 95\% coverage probability, the average length of a 95\%
confidence interval, and the average of their plug-in estimator for the
asymptotic variance. The confidence intervals are based on the normal
approximation, with the asymptotic variances being estimated by their sample
analogues. We also compute the semiparametric efficiency bound under each
design to allow one to assess the potential loss of efficiency/accuracy
associated with using inefficient DID estimators for the ATT.

\subsection{Simulation 1: panel data are available}

We first discuss the case where panel data are available. For a generic $%
W=\left( W_{1},W_{2},W_{3},W_{4}\right) ^{\prime },$ let%
\begin{eqnarray*}
f_{reg}\left( W\right) &=&210+27.4\cdot W_{1}+13.7\cdot \left(
W_{2}+W_{3}+W_{4}\right) , \\
f_{ps}\left( W\right) &=&0.75\cdot (-W_{1}+0.5\cdot W_{2}-0.25\cdot
W_{3}-0.1\cdot W_{4}).
\end{eqnarray*}%
Let $\mathbf{X}=\left( X_{1},X_{2},\right. $ $\left. X_{3},X_{4}\right)
^{\prime }$ be distributed as $N\left( 0,I_{4}\right) $, and $I_{4}$ be the $%
4\times 4$ identity matrix. For $j=1,2,3,4$, let $Z_{j}=\left. \left( \tilde{%
Z}-\mathbb{E}\left[ \tilde{Z}\right] \right) \right/ \sqrt{Var\left( \tilde{Z%
}\right) }$, where $\tilde{Z}_{1}=\exp \left( 0.5X_{1}\right) $, $\tilde{Z}%
_{2}=10+X_{2}/\left( 1+\exp \left( X_{1}\right) \right) $, $\tilde{Z}%
_{3}=\left( 0.6+\left. X_{1}X_{3}\right/ 25\right) ^{3}$ and $\tilde{Z}%
_{4}=\left( 20+X_{2}+X_{4}\right) ^{2}$.

Building on \cite{Kang2007}, we consider the following data generating
processes (DGPs):%
\begin{align*}
DGP1.~~~\ Y_{0}\left( 0\right) & =f_{reg}\left( Z\right) +v\left( Z,D\right)
+\varepsilon _{0},~~~~~~~~~~~~~~~Y_{1}\left( d\right) =2\cdot f_{reg}\left(
Z\right) +v\left( Z,D\right) +\varepsilon _{1}\left( d\right) ,~d=0,1, \\
p\left( Z\right) & =\frac{\exp \left( f_{ps}\left( Z\right) \right) }{1+\exp
\left( f_{ps}\left( Z\right) \right) }\text{, }~~~~~~~~~~~~~~~~~~~~~\text{~~~%
}D=1\left\{ p\left( Z\right) \geq U\right\} ; \\
DGP2.~~~\ Y_{0}\left( 0\right) & =f_{reg}\left( Z\right) +v\left( Z,D\right)
+\varepsilon _{0},~~~~~~~~~~~~~~~Y_{1}\left( d\right) =2\cdot f_{reg}\left(
Z\right) +v\left( Z,D\right) +\varepsilon _{1}\left( d\right) ,~d=0,1, \\%
[1ex]
p\left( X\right) & =\frac{\exp \left( f_{ps}\left( X\right) \right) }{1+\exp
\left( f_{ps}\left( X\right) \right) }\text{, }~~~~~~~~~~~~~~~~~~~~~\text{~~~%
}D=1\left\{ p\left( X\right) \geq U\right\} ; \\
DGP3.~~~\ Y_{0}\left( 0\right) & =f_{reg}\left( X\right) +v\left( X,D\right)
+\varepsilon _{0},~~~~~~~~~~~~~~~Y_{1}\left( d\right) =2\cdot f_{reg}\left(
X\right) +v\left( X,D\right) +\varepsilon _{1}\left( d\right) ,~d=0,1, \\
p\left( Z\right) & =\frac{\exp \left( f_{ps}\left( Z\right) \right) }{1+\exp
\left( f_{ps}\left( Z\right) \right) }\text{, }~~~~~~~~~~~~~~~~~~~~~\text{~~~%
}D=1\left\{ p\left( Z\right) \geq U\right\} ; \\
DGP4.~~~\ Y_{0}\left( 0\right) & =f_{reg}\left( X\right) +v\left( X,D\right)
+\varepsilon _{0},~~~~~~~~~~~~~~~Y_{1}\left( d\right) =2\cdot f_{reg}\left(
X\right) +v\left( X,D\right) +\varepsilon _{1}\left( d\right) ,~d=0,1, \\
p\left( X\right) & =\frac{\exp \left( f_{ps}\left( X\right) \right) }{1+\exp
\left( f_{ps}\left( X\right) \right) }\text{, }~~~~~~~~~~~~~~~~~~~~~\text{~~~%
}D=1\left\{ p\left( X\right) \geq U\right\} ,
\end{align*}%
where $\varepsilon _{0}$, $\varepsilon _{1}\left( d\right) $, $d=0,1$ are
independent standard normal random variables, $U$ is an independent standard
uniform random variable, and for a generic $W$, $v\left( W,D\right) $ is an
independent normal random variable with mean $D\cdot f_{reg}\left( W\right) $
and variance one. The available data are $\left\{
Y_{0,i},Y_{1,i},D_{i},Z_{i}\right\} _{i=1}^{n}$, where $Y_{0}=Y_{0}\left(
0\right) $, and $Y_{1}=DY_{1}\left( 1\right) +\left( 1-D\right) Y_{1}\left(
0\right) $. In the aforementioned DGPs, the true ATT is zero, and $v$ plays
the role of time-invariant unobserved heterogeneity.

Given that we focus on the empirically relevant setting where the observed
covariates $Z$ enter all working models linearly, it is clear that in $DPG1$%
, both propensity score (PS) and OR working models are correctly specified.
In $DGP2$, only the OR working model is correctly specified, whereas in $%
DGP3 $ only the PS working model is correctly specified. In $DGP4$, all
working models are misspecified. The simulation results are presented in
Table \ref{tab:MCpanel}. \medskip

\begin{table}[h]
\caption{Monte Carlo results under designs $DGP1-DGP4$ with panel data.
Sample size $n=1,000$.}
\label{tab:MCpanel}\centering%
\begin{adjustbox}{ max width=1\linewidth, max totalheight=1\textheight, keepaspectratio}
\begin{threeparttable}
    \begin{tabular}{@{}lccccccccccccc@{}} \toprule 
      \noalign{\vskip 2mm}    & \multicolumn{6}{c}{DGP1: OR correct, PS correct} &   & \multicolumn{6}{c}{DGP2: OR correct, PS incorrect} \\ 
          & \multicolumn{6}{c}{Semiparametric Efficiency Bound: 11.1 } &   & \multicolumn{6}{c}{Semiparametric Efficiency Bound: 11.6}  \\ \cline{2-7} \cline{9-14}
        \noalign{\vskip 1mm}  & \multicolumn{1}{c}{Av. Bias} &\multicolumn{1}{c}{Med. Bias} & \multicolumn{1}{c}{RMSE} & \multicolumn{1}{c}{Asy. V} & \multicolumn{1}{c}{Cover} & \multicolumn{1}{c}{CIL} & \multicolumn{1}{c}{} & \multicolumn{1}{c}{Av. Bias} &\multicolumn{1}{c}{Med. Bias} &  \multicolumn{1}{c}{RMSE} & \multicolumn{1}{c}{Asy. V} & \multicolumn{1}{c}{Cover} & \multicolumn{1}{c}{CIL} \\ \cline{2-7} \cline{9-14}

    $\widehat{\tau}^{fe}$ &-20.952 & -20.965 & 21.123 & 6392.2 & 0.000 & 9.906 &   & -19.286 & -19.287 & 19.468 & 6640.3 & 0.000 & 10.095 \\
    $\widehat{\tau}^{reg}$  & -0.001 & -0.001 & 0.100 & 10.2 & 0.950 & 0.396 &   & -0.001 & -0.001 & 0.100 & 10.1 & 0.949 & 0.394 \\
           $\widehat{\tau}^{ipw,p}$  & 0.026 & 0.195 & 2.774 & 8078.0 & 0.952 & 10.441 &   & 2.010 & 2.054 & 3.298 & 7048.3 & 0.838 & 9.819 \\
      $\widehat{\tau}^{ipw,p}_{std}$ & 0.008 & -0.013 & 1.132 & 1286.4 & 0.948 & 4.309 &   & -0.794 & -0.798 & 1.225 & 891.7 & 0.856 & 3.623 \\
    $\widehat{\tau}^{dr,p}$ & -0.001 & 0.000 & 0.106 & 11.1 & 0.947 & 0.412 &   & -0.001 & -0.002 & 0.104 & 10.7 & 0.947 & 0.404 \\
     $\widehat{\tau}^{dr,p}_{imp}$ & -0.001 & 0.000 & 0.106 & 10.9 & 0.945 & 0.409 &   & -0.001 & -0.001 & 0.104 & 10.6 & 0.945 & 0.404 \\
    
            \noalign{\vskip 4mm}  & \multicolumn{6}{c}{DGP3: OR incorrect, PS correct} &   & \multicolumn{6}{c}{DGP4: OR incorrect, PS incorrect} \\   
         & \multicolumn{6}{c}{Semiparametric Efficiency Bound: 11.1 } &   & \multicolumn{6}{c}{Semiparametric Efficiency Bound: 11.6}  \\ \cline{2-7} \cline{9-14}
        \noalign{\vskip 1mm}  & \multicolumn{1}{c}{Av. Bias} &\multicolumn{1}{c}{Med. Bias} & \multicolumn{1}{c}{RMSE} & \multicolumn{1}{c}{Asy. V} & \multicolumn{1}{c}{Cover} & \multicolumn{1}{c}{CIL} & \multicolumn{1}{c}{} & \multicolumn{1}{c}{Av. Bias} &\multicolumn{1}{c}{Med. Bias} &  \multicolumn{1}{c}{RMSE} & \multicolumn{1}{c}{Asy. V} & \multicolumn{1}{c}{Cover} & \multicolumn{1}{c}{CIL} \\ \cline{2-7} \cline{9-14}
        
             $\widehat{\tau}^{fe}$ & -13.170 & -13.194 & 13.364 & 12687.9 & 0.004 & 13.960 &   & -16.385 & -16.393 & 16.538 & 13160.7 & 0.000 & 14.217 \\
     $\widehat{\tau}^{reg}$  & -1.384 & -1.365 & 1.868 & 1514.4 & 0.800 & 4.816 &   & -5.204 & -5.171 & 5.364 & 1666.6 & 0.015 & 5.053 \\
    $\widehat{\tau}^{ipw,p}$ & 0.011 & 0.158 & 3.198 & 10062.5 & 0.947 & 11.777 &   & -1.085 & -1.017 & 2.656 & 6151.4 & 0.949 & 9.308 \\
   $\widehat{\tau}^{ipw,p}_{std}$ & -0.030 & -0.032 & 1.427 & 1988.0 & 0.945 & 5.484 &   & -3.954 & -3.949 & 4.215 & 2156.5 & 0.228 & 5.717 \\
     $\widehat{\tau}^{dr,p}$  & -0.051 & -0.046 & 1.214 & 1400.9 & 0.942 & 4.613 &   & -3.188 & -3.183 & 3.454 & 1704.9 & 0.308 & 5.075 \\
   $\widehat{\tau}^{dr,p}_{imp}$& -0.071 & -0.064 & 1.015 & 971.2 & 0.942 & 3.858 &   & -2.529 & -2.514 & 2.720 & 970.1 & 0.274 & 3.856 \\
  
    \bottomrule
\end{tabular}%
\begin{tablenotes}[para,flushleft]
\small{
Notes: Simulations based on 10,000 Monte Carlo experiments. $\widehat{\tau}^{fe}$ is the TWFE outcome regression estimator of $\tau^{fe}$ in (\ref{fe}),  $\widehat{\tau}^{reg}$ is the OR-DID estimator (\ref{att.reg1}), $\widehat{\tau}^{dr,p}$ is the IPW DID estimator (\ref{att.ipw}), 
$\widehat{\tau}^{ipw,p}_{std}$ is the standardized IPW DID estimator (\ref{att.ipw2}), $\widehat{\tau}^{dr,p}$ is our proposed DR DID estimator (\ref{dr.p.estimator}), and  $\widehat{\tau}^{dr,p}_{imp}$ is our proposed DR DID estimator (\ref{dr.p.imp.estimator}). We use a linear OR working model and a logistic PS working model, where the unknown parameters are estimated via OLS and maximum likelihood, respectively, except for $\widehat{\tau}^{dr,p}_{imp}$, where we use the estimation methods described in Section \ref{sec:firststep}. Finally, ``Av. Bias'', ``Med. Bias'', ``RMSE'', ``Asy. V'', ``Cover'' and ``CIL', stand for the average simulated bias, median simulated bias, simulated root mean-squared errors, average of the plug-in estimator for the asymptotic variance, 95\% coverage probability, and 
95\% confidence interval length, respectively. See the main text for further details.}
\end{tablenotes}
\end{threeparttable}
\end{adjustbox}
\end{table}

First, note that the TWFE estimator $\widehat{\tau }^{fe}$ is severely
biased and its confidence interval for the ATT has almost zero coverage in
all analyzed DGPs. These results should not be unexpected, because, as
discussed in Remark \ref{rem:twfe}, $\widehat{\tau }^{fe}$ implicitly rules
out covariate-specific trends, and when these are relevant, like in the
considered DGPs, the estimand associated with $\widehat{\tau }^{fe}$ is not
the ATT. As so, policy evaluations based on $\widehat{\tau }^{fe}$ can be
misleading.

The results in Table \ref{tab:MCpanel} also suggest that, when both the OR
and PS working models are correctly specified, all semiparametric estimators
for the ATT show little to no Monte Carlo bias, but $\widehat{\tau }^{reg}$, 
$\widehat{\tau }^{dr,p}$ and $\widehat{\tau }_{imp}^{dr,p}$ dominate the IPW
DID estimators $\widehat{\tau }^{ipw,p}$ and $\widehat{\tau }_{std}^{ipw,p}$%
on the basis of bias, root mean square error, asymptotic variance, and
length of the confidence interval. Indeed, both IPW DID estimator seem to be
substantially less efficient than $\widehat{\tau }^{reg}$, $\widehat{\tau }%
^{dr,p}$ and $\widehat{\tau }_{imp}^{dr,p}$. The performance of these last
three estimators are very close, though $\widehat{\tau }^{reg}$ tends to be
more efficient than the other two DR DID estimators. Given that $\widehat{%
\tau }^{reg}$ exploits additional assumptions when compared to $\widehat{%
\tau }^{dr,p}$ and $\widehat{\tau }_{imp}^{dr,p}$, such a result is not
unexpected. Also note that the \cite{Hajek1971} type IPW estimator $\widehat{%
\tau }_{std}^{ipw,rc}$ is more stable than the \cite{Horvitz1952} type IPW
estimator $\widehat{\tau }^{ipw,rc}$: the RMSE (the asymptotic variance) of $%
\widehat{\tau }^{ipw,rc}$ are more than two (four) times bigger than that of 
$\widehat{\tau }_{std}^{ipw,rc}$. Such a finding highlights the practical
importance of using weights that are normalized to sum up to one.

When only the OR working model is correctly specified, our proposed DR DID
estimators $\widehat{\tau }^{dr,p}$ and $\widehat{\tau }_{imp}^{dr,p}$ are
competitive with the OR DID estimator $\widehat{\tau }^{reg}$, while the IPW
DID estimators are biased, as one should expect. On the other hand, when
only the PS working model is correctly specified, the IPW and DR estimators
show little to no bias, while $\widehat{\tau }^{reg}$ displays
non-negligible bias. Here, it is worth emphasizing that $\widehat{\tau }%
^{dr,p}$ and $\widehat{\tau }_{imp}^{dr,p}$ drastically outperform $\widehat{%
\tau }^{ipw,p}$ and $\widehat{\tau }_{std}^{ipw,p}$, with $\widehat{\tau }%
_{imp}^{dr,p}$ also showing substantial improvements with respect to both $%
\widehat{\tau }^{dr,p}$ and $\widehat{\tau }_{std}^{ipw,p}$. When one
compares the two IPW estimators, the role played by the normalized weights
is again clear, as $\widehat{\tau }_{std}^{ipw,p}$ is again much more
\textquotedblleft stable\textquotedblright\ than $\widehat{\tau }^{ipw,p}$.

When both OR and PS working models are misspecified, not unexpectedly all
estimators have non-negligible biases and inference procedures are, in
general, misleading. In this scenario, our DR DID estimators have smaller
biases and RMSE than the OR and the normalized IPW DID estimators, with $%
\widehat{\tau }_{imp}^{dr,p}$ strictly dominating $\widehat{\tau }^{dr,p}.$
However, the \cite{Horvitz1952} IPW DID estimator $\widehat{\tau }^{ipw,p}$
seems to perform best in this DGP.

In terms of efficiency, the results in Table \ref{tab:MCpanel} show that the
estimated asymptotic variance of $\widehat{\tau }^{dr,p}$ and $\widehat{\tau 
}_{imp}^{dr,p}$ are very close to the semiparametric efficiency bound when
both the PS and OR regression are correctly specified, which is in agreement
with our locally efficiency results in Theorems \ref{prop:att.br.panel} and %
\ref{th:att.panel} (in the Appendix). When the PS is misspecified but the OR
is not, the estimated asymptotic variances of $\widehat{\tau }^{dr,p}$ and $%
\widehat{\tau }_{imp}^{dr,p}$ are still close to the semiparametric
efficiency bound in this particular DGP, though we emphasize that this is
not predicted by our results and can be a feature of this particular DGP.
Finally, we note that when the OR is misspecified but the PS is not, the
estimated asymptotic variances of our proposed DR DID estimators $\widehat{%
\tau }^{dr,p}$ and $\widehat{\tau }_{imp}^{dr,p}$ are far from the
semiparametric efficiency bound, with $\widehat{\tau }_{imp}^{dr,p}$
outperforming $\widehat{\tau }^{dr,p}$ in terms of efficiency in this
particular DGP.

\subsection{Simulation 2: repeated cross-section data are available}

We now analyze the performance of the DID estimators for the ATT when one
only observes repeated cross-section data. To do so, we consider the same
DGPs as in the panel data framework, but instead of observing data on $%
\left( Y_{0},Y_{1},D,Z\right) ,$ one observes data on $\left(
Y_{0},D,Z\right) $ if $T=0$, or on $\left( Y_{1},D,Z\right) $ if $T=1$,
where $T=1\left\{ U_{T}\leq \lambda \right\} $, and $U_{T}$ is a standard
uniform random variable, and $\lambda \in \left( 0,1\right) $ a fixed
constant.

\begin{table}[h]
\caption{Monte Carlo results under designs $DGP1-DGP4$ with repeated cross
section data. Sample size $n=1,000$, and $\protect\lambda =0.5$.}
\label{tab:MCrc}\centering%
\begin{adjustbox}{ max width=1\linewidth, max totalheight=1\textheight, keepaspectratio}
\begin{threeparttable}
    \begin{tabular}{@{}lccccccccccccc@{}} \toprule 
       \noalign{\vskip 2mm}    & \multicolumn{6}{c}{DGP1: OR correct, PS correct} &   & \multicolumn{6}{c}{DGP2: OR correct, PS incorrect} \\ 
          & \multicolumn{6}{c}{Semiparametric Efficiency Bound: 44.4 } &   & \multicolumn{6}{c}{Semiparametric Efficiency Bound: 46.4}  \\ \cline{2-7} \cline{9-14}
        \noalign{\vskip 1mm}  & \multicolumn{1}{c}{Av. Bias} &\multicolumn{1}{c}{Med. Bias} & \multicolumn{1}{c}{RMSE} & \multicolumn{1}{c}{Asy. V} & \multicolumn{1}{c}{Cover} & \multicolumn{1}{c}{CIL} & \multicolumn{1}{c}{} & \multicolumn{1}{c}{Av. Bias} &\multicolumn{1}{c}{Med. Bias} &  \multicolumn{1}{c}{RMSE} & \multicolumn{1}{c}{Asy. V} & \multicolumn{1}{c}{Cover} & \multicolumn{1}{c}{CIL} \\ \cline{2-7} \cline{9-14}

         $\widehat{\tau}^{fe}$  & -20.792 & -20.741 & 21.099 & 12773.9 & 0.000 & 13.996 &   & -19.178 & -19.125 & 19.529 & 13240.7 & 0.001 & 14.247 \\
      $\widehat{\tau}^{reg}$   & 0.026 & -0.030 & 7.588 & 57417.8 & 0.951 & 29.675 &   & -0.024 & -0.057 & 8.191 & 66577.5 & 0.948 & 31.945 \\
     $\widehat{\tau}^{ipw,rc}$ & -0.662 & -0.932 & 55.971 & 3090077.6 & 0.949 & 217.762 &   & 1.820 & 1.506 & 55.050 & 3023548.1 & 0.949 & 215.449 \\
    $\widehat{\tau}^{ipw,rc}_{std}$  & -0.050 & -0.125 & 9.648 & 92235.7 & 0.949 & 37.560 &   & -0.812 & -0.698 & 9.814 & 94343.0 & 0.946 & 38.031 \\
   $\widehat{\tau}^{dr,rc}_{1}$ & 0.013 & -0.007 & 3.041 & 9222.0 & 0.950 & 11.893 &   & -0.010 & -0.022 & 3.281 & 10686.4 & 0.949 & 12.799 \\
   $\widehat{\tau}^{dr,rc}_{2}$ & 0.004 & 0.003 & 0.216 & 44.4 & 0.944 & 0.824 &   & 0.000 & 0.001 & 0.211 & 42.3 & 0.945 & 0.805 \\
   $\widehat{\tau}^{dr,rc}_{1,imp}$ & 0.014 & -0.008 & 3.041 & 9220.1 & 0.951 & 11.892 &   & -0.009 & -0.022 & 3.282 & 10686.2 & 0.949 & 12.799 \\
      $\widehat{\tau}^{dr,rc}_{2,imp}$ & 0.005 & 0.002 & 0.216 & 42.1 & 0.937 & 0.803 &   & 0.000 & 0.001 & 0.213 & 41.3 & 0.940 & 0.796 \\

           \noalign{\vskip 4mm}  & \multicolumn{6}{c}{DGP3: OR incorrect, PS correct} &   & \multicolumn{6}{c}{DGP4: OR incorrect, PS incorrect} \\   
         & \multicolumn{6}{c}{Semiparametric Efficiency Bound: 44.4 } &   & \multicolumn{6}{c}{Semiparametric Efficiency Bound: 46.4}  \\ \cline{2-7} \cline{9-14}
        \noalign{\vskip 1mm}  & \multicolumn{1}{c}{Av. Bias} &\multicolumn{1}{c}{Med. Bias} & \multicolumn{1}{c}{RMSE} & \multicolumn{1}{c}{Asy. V} & \multicolumn{1}{c}{Cover} & \multicolumn{1}{c}{CIL} & \multicolumn{1}{c}{} & \multicolumn{1}{c}{Av. Bias} &\multicolumn{1}{c}{Med. Bias} &  \multicolumn{1}{c}{RMSE} & \multicolumn{1}{c}{Asy. V} & \multicolumn{1}{c}{Cover} & \multicolumn{1}{c}{CIL} \\ \cline{2-7} \cline{9-14}

  $\widehat{\tau}^{fe}$ & -13.131 & -13.092 & 14.058 & 25446.9 & 0.260 & 19.766 &   & -16.330 & -16.354 & 17.126 & 26347.3 & 0.114 & 20.112 \\
    $\widehat{\tau}^{reg}$  & -1.376 & -1.397 & 8.137 & 64143.7 & 0.942 & 31.378 &   & -5.338 & -5.437 & 9.977 & 72665.8 & 0.908 & 33.397 \\
    $\widehat{\tau}^{ipw,rc}$ & -0.973 & -1.452 & 57.262 & 3241967.3 & 0.947 & 223.050 &   & -1.391 & -0.980 & 55.178 & 3101777.5 & 0.952 & 218.233 \\
   $\widehat{\tau}^{ipw,rc}_{std}$  & 0.051 & -0.011 & 9.428 & 86806.4 & 0.943 & 36.483 &   & -4.149 & -4.387 & 10.520 & 94034.1 & 0.930 & 37.971 \\
    $\widehat{\tau}^{dr,rc}_{1}$ & -0.086 & -0.083 & 5.692 & 31830.9 & 0.945 & 22.060 &   & -3.342 & -3.375 & 7.071 & 38663.1 & 0.916 & 24.290 \\
    $\widehat{\tau}^{dr,rc}_{2}$ & -0.029 & -0.022 & 4.742 & 21869.3 & 0.942 & 18.261 &   & -3.275 & -3.249 & 6.016 & 24194.2 & 0.886 & 19.159 \\
   $\widehat{\tau}^{dr,rc}_{1,imp}$  & -0.119 & -0.102 & 4.837 & 23038.9 & 0.945 & 18.804 &   & -2.689 & -2.708 & 5.564 & 23473.3 & 0.913 & 18.979 \\
    $\widehat{\tau}^{dr,rc}_{2,imp}$ & -0.076 & -0.081 & 4.062 & 15765.2 & 0.944 & 15.550 &   & -2.614 & -2.610 & 4.845 & 15769.1 & 0.892 & 15.552 \\

    \bottomrule
\end{tabular}%
\begin{tablenotes}[para,flushleft]
\small{
Notes: Simulations based on 10,000 Monte Carlo experiments. $\widehat{\tau}^{fe}$ is the TWFE outcome regression estimator of $\tau^{fe}$ in (\ref{fe}),  $\widehat{\tau}^{reg}$ is the OR-DID estimator (\ref{att.reg1}), $\widehat{\tau}^{dr,rc}$ is the IPW DID estimator based on the sample
analogue of (\ref{abadie.rc}), $\widehat{\tau}^{ipw,rc}_{std}$ is the standardized IPW DID estimator (\ref{att.ipw2.rc}), and $\widehat{\tau}_{1}^{dr,rc},$ $\widehat{\tau }_{2}^{dr,rc}$, $\widehat{\tau }%
_{1,imp}^{dr,rc}$ and $\widehat{\tau }_{2,imp}^{dr,rc}$are our proposed DR DID estimators given in (\ref{dr.rc1.estimator}), (\ref{dr.rc2.estimator}), and in (\ref{dr.rc.imp1.estimator}) and (\ref{dr.rc.imp2.estimator}) in Section \ref{App:improved.rc}. We use a linear OR working model and a logistic PS working model, where the unknown parameters are estimated via OLS and maximum likelihood, respectively, except for $\widehat{\tau}^{dr,rc}_{1,imp}$ and  $\widehat{\tau}^{dr,rc}_{2,imp}$ where we use the estimation methods described in Section \ref{App:improved.rc}.
 Finally, ``Av. Bias'',``Med. Bias'',  ``RMSE'', ``Asy. V'', ``Cover'' and ``CIL', stand for the average simulated bias, median simulated bias, simulated root mean-squared errors, , average of the plug-in estimator for the asymptotic variance, 95\% coverage probability, and 
95\% confidence interval length, respectively. See the main text for further details.}
\end{tablenotes}
\end{threeparttable}
\end{adjustbox}
\end{table}

Table \ref{tab:MCrc} presents the simulation results with $\lambda =0.5$ and
with $n\equiv n_{1}+n_{0}=1,000$.\footnote{%
Simulation results with $\lambda =0.25$ and $\lambda =0.75$ reached
analogous conclusions to those discussed below and are available upon
request.} Overall, the simulation exercise reveals that the efficiency
bound, RMSE, asymptotic variance, and confidence interval length of the
considered DID estimators are much larger when only repeated cross-section
data are available than when panel data are available. In light of Corollary %
\ref{cor:var.bound}, such a result should be expected, though the magnitude
of such loss of efficiency can be striking. In addition, the results in
Table \ref{tab:MCrc} reveal that: $\left( i\right) $ the TWFE estimator $%
\widehat{\tau }^{fe}$ is severely biased for the ATT in all DGPs, just like
in the panel data case; $\left( ii\right) $ the IPW estimator with
standardized weights $\widehat{\tau }_{std}^{ipw,rc}$ is much more stable
and efficient than $\widehat{\tau }^{ipw,rc}$ in all DGPs, and, as one
should expect, when the PS working model is misspecified, these IPW
estimators display non-negligible biases; $\left( iii\right) $ as one should
expect, the OR DID estimator displays non-negligible bias when the OR
working models are misspecified; $\left( iv\right) $ all four DR DID
estimators display little to no bias when one of the working models is
correctly specified, but the locally efficient DR DID estimators $\widehat{%
\tau }_{2}^{dr,rc}$ and $\widehat{\tau }_{2,imp}^{dr,rc}$ present important
efficiency gains when compared to all other DID estimators, including $%
\widehat{\tau }_{1}^{dr,rc}$ and $\widehat{\tau }_{1,imp}^{dr,rc}$. These
gains in efficiency are more pronounced when the OR models are correctly
specified. The simulation results also show that $\left( v\right) $ when one
compares the performance of the further improved DR DID estimators $\widehat{%
\tau }_{1,imp}^{dr,rc}$ and $\widehat{\tau }_{2,imp}^{dr,rc}$ with the
\textquotedblleft traditional\textquotedblright\ DR DID estimators $\widehat{%
\tau }_{1}^{dr,rc}$ and $\widehat{\tau }_{2}^{dr,rc}$, it is clear that
appropriately choosing the estimation methods for the nuisance parameters
can have practical consequences, especially when the outcome regression
working models are misspecified.

In terms of efficiency, the results in Table \ref{tab:MCrc} highlight that,
when all working models are correctly specified, the estimated asymptotic
variances of $\widehat{\tau }_{2}^{dr,rc}$ and $\widehat{\tau }%
_{2,imp}^{dr,rc}$ are indeed close to the semiparametric efficiency bound,
but the asymptotic variances $\widehat{\tau }_{1}^{dr,rc}$ and $\widehat{%
\tau }_{1,imp}^{dr,rc}$ are substantially higher than the semiparametric
efficiency bound; these findings are in agreement with our locally
efficiency results in Theorems \ref{prop:att.br.rc} and \ref{th:att.rc} (in
the Appendix). Similarly to the panel data case, we find that, in this
specific DGP, the estimated asymptotic variances of $\widehat{\tau }%
_{2}^{dr,rc}$ and $\widehat{\tau }_{2,imp}^{dr,rc}$ are still close to the
semiparametric efficiency bound when the outcome regressions are correctly
specified but the PS is not, but not when the PS is correctly specified but
the outcome regressions are not.

\section{Empirical illustration: the effect of job training on earnings 
\label{sec:applications}}

In a very influential study, \cite{Lalonde1986} analyzes whether different
treatment effect estimators based on observational data are able to
replicate the experimental findings of the NSW job training program on post
treatment earnings. His negative results led to an increased awareness of
the potential pitfalls of observational data and helped spur the use of
randomized controlled trials among economists. In addition, alternative
policy evaluation tools arose to overcome \textquotedblleft LaLonde's
critique\textquotedblright\ of observational estimators. Two prominent
examples are the propensity score matching (PSM), see e.g.~\cite%
{Dehejia1999, Dehejia2002} (henceforth DW) and the difference-in-differences
matching, see e.g.~\cite{Heckman1997} and \cite{Smith2005} (henceforth ST).
For instance, DW show that PSM can replicate the experimental benchmark of
the NSW for a particular subsample of the original data. ST, on the other
hand, cast doubt on the \textquotedblleft
generalizability\textquotedblright\ of DW PSM results to a larger population
and argue that the conclusions may be sensitive to the propensity score
specification. ST also argue that for the NSW data,
difference-in-differences matching estimators may be more suitable than
cross-section PSM, as they can account for time-invariant unobserved
confounding factors.

Motivated by ST findings, in what follows, we focus on DID estimators and
evaluate whether our proposed DR DID estimators can better reduce the
selection bias when compared to other DID estimation procedures. We analyze
three different experimental samples --- the original LaLonde experimental
sample, the DW sample, and the \textquotedblleft early random
assignment\textquotedblright\ (early RA) subsample of the DW sample
considered by ST --- and consider data from the Current Population Survey
(CPS) to form a non-experimental comparison group. The pre-treatment
covariates in the data include age, years of education, real earnings in
1974, and dummy variables for high school dropout, married, black, and
Hispanic. The outcome of interest is real earnings in 1978. We also observe
real earnings in 1975, which we use as the pre-treatment outcome $Y_{0}$.
The experimental benchmark for the ATT is equal to $\$886$ (s.e. $\$488$), $%
\$1794$ (s.e. $\$671$), and $\$2748$ (s.e. $\$1005$) for the LaLonde, DW,
and early RA sample, respectively. For additional description and summary
statistics for each sample, see \cite{Smith2005}.

Following ST, we focus on estimating the average \textquotedblleft
evaluation bias\textquotedblright\ of different DID estimators. This is only
made possible given the availability of experimental data. First,
randomization ensures that both \textquotedblleft
treatment\textquotedblright\ groups are comparable in terms of self
selection. Second, given that randomized-out individuals did not receive
training via NSW, the impact of NSW is known to be zero in this group. Thus,
applying different DID estimators to data from randomized-out individuals
(our pseudo treated group in this exercise) and nonexperimental CPS
comparison observations (our comparison group in this exercise) should
produce an estimated ATT equal to zero, if these DID estimators are
consistent. Deviations from zero are what we call evaluation bias.\footnote{%
An alternative way to estimate \textquotedblleft evaluation
bias\textquotedblright\ is to compare the ATT using the experimental data
with ATT using data from randomized-in and nonexperimental comparison units.
This is the approach taken by \cite{Lalonde1986} and \cite{Dehejia1999,
Dehejia2002}. A disadvantage of this approach compared to the one we and 
\cite{Smith2005} use is that experimental ATT estimates are also random and
may differ from the \textquotedblleft true\textquotedblright\ ATT. Thus, the
computation of \textquotedblleft true\textquotedblright\ evaluation biases
is much more challenging if not impossible. In any case, results treating
the experimental ATT as true effects lead to similar conclusions and are
available upon request.}

Like in the Monte Carlo simulation exercises, we compare our proposed DR DID
estimators $\widehat{\tau }^{dr,p}$ and $\widehat{\tau }_{imp}^{dr,p}$ with
the TWFE estimator $\widehat{\tau }^{fe}$ based on (\ref{fe}), the OR DID
estimator $\widehat{\tau }^{reg}$ as defined in (\ref{att.reg1}), and the 
\cite{Horvitz1952} type IPW DID estimator proposed by \cite{Abadie2005}, $%
\widehat{\tau }^{ipw,p}$, as defined in (\ref{att.ipw}). We also consider
the \cite{Hajek1971} type IPW estimator $\widehat{\tau }_{std}^{ipw,p}$ as
defined in (\ref{att.ipw2}). We assume that the outcome models are linear in
parameters and that the propensity score follows a logistic specification.
The unknown parameters are estimated using ordinary least squares (OLS) and
maximum likelihood, respectively, except in $\widehat{\tau }_{imp}^{dr,p}$,
where we use the estimation methods described in Section \ref{sec:firststep}.

In order to assess the sensitivity of the findings with respect to the model
specifications, we consider three different specifications for how
covariates enter into each model: $\left( i\right) $ a linear specification
where all covariates enter the models linearly; $\left( ii\right) $ a
specification in the spirit of DW, which adds to the linear specification a
dummy for zero earnings in 1974, age squared, age cubed divided by $1000$,
years of schooling squared, and an interaction term between years of
schooling and real earnings in 1974; and $\left( iii\right) $ an
\textquotedblleft augmented DW\textquotedblright\ specification, which adds
to the \textquotedblleft DW\textquotedblright\ specification the
interactions between married and real earnings in 1974, and between married
and zero earnings in 1974 --- these two interaction terms were used in \cite%
{Firpo2007}.

Table \ref{tab:st-bias} summarizes the results. Standard errors are reported
in parentheses and the estimated evaluation biases relative to the
experimental ATT benchmark are reported in brackets. As argued by ST, these
\textquotedblleft relative biases\textquotedblright\ are useful for
comparing DID estimators within each sample, but as the experimental
benchmark estimates for the ATT vary substantially among the three
experimental samples, they should not be used for comparing DID estimators
across samples. 
\afterpage{
\begin{landscape}
\vspace*{\fill}
\begin{table}[h]
\caption{Evaluation bias of different difference-in-differences estimators
for the effect of of training on real earnings in 1978. NSW data with CPS
comparison group. }\centering\centering\begin{adjustbox}{ max width=1\linewidth, max totalheight=1\textheight, keepaspectratio}
\begin{threeparttable}

     \begin{tabular}{lcccccccccccccccccccc@{}} \toprule
    \noalign{\vskip 2mm}  \multicolumn{1}{l}{ }                               & \multicolumn{6}{c}{Results for Lalonde sample}                                                                                                    &     \phantom{ab}                               & \multicolumn{6}{c}{Results for DW sample}                                                                                                         &                 \phantom{ab}                   & \multicolumn{6}{c}{Results for Early RA sample} \\ 
 
\noalign{\vskip -4mm}  \\
   \multicolumn{1}{l}{}                                & \multicolumn{6}{c}{Evaluation Bias: ATT$=0$}                                                                                                             &                                & \multicolumn{6}{c}{Evaluation Bias: ATT$=0$}                                                                                                             &                                    & \multicolumn{6}{c}{Evaluation Bias: ATT$=0$} \\ \noalign{\vskip 2mm} \cline{2-7} \cline{9-14} \cline{16-21}  
\noalign{\vskip 2mm}    Spec.                                 & \multicolumn{1}{c}{$\widehat{\tau}^{dr,p}$} & \multicolumn{1}{c}{$\widehat{\tau}^{dr,p}_{imp}$}            & \multicolumn{1}{c}{$\widehat{\tau}^{reg}$}            & \multicolumn{1}{c}{  $\widehat{\tau}^{ipw,p}$}         &
\multicolumn{1}{c}{  $\widehat{\tau}^{ipw,p}_{std}$}         &  \multicolumn{1}{c}{  $\widehat{\tau}^{fe}$}    &                                    & \multicolumn{1}{c}{$\widehat{\tau}^{dr,p}$} & \multicolumn{1}{c}{$\widehat{\tau}^{dr,p}_{imp}$}                    & \multicolumn{1}{c}{  $\widehat{\tau}^{reg}$}            & \multicolumn{1}{c}{  $\widehat{\tau}^{ipw,p}$}      &     
 \multicolumn{1}{c}{  $\widehat{\tau}^{ipw,p}_{std}$}         & 
  \multicolumn{1}{c}{  $\widehat{\tau}^{fe}$}   &                                    & \multicolumn{1}{c}{$\widehat{\tau}^{dr,p}$}      & \multicolumn{1}{c}{$\widehat{\tau}^{dr,p}_{imp}$}         & \multicolumn{1}{c}{$\widehat{\tau}^{reg}$}            & \multicolumn{1}{c}{  $\widehat{\tau}^{ipw,p}$}  &
  \multicolumn{1}{c}{  $\widehat{\tau}^{ipw,p}_{std}$}         &  \multicolumn{1}{c}{  $\widehat{\tau}^{fe}$}      \\ \noalign{\vskip 1mm}  \midrule

     \multicolumn{1}{l}{Lin.}         & -871 & -901 & -1301 & -1108 &-1022 & 868 &   & 253 & 253 & -230 & 188 & 155 & 2092 &   & -434 & -441 & -831 & -516 &-515 &1136 \\
                                      \noalign{\vskip -1mm}  &        (396) & (394) & (350) & (409) &(398) & (353) &   & (451) & (452) & (408) & (459) &(452) & (459) &   & (605) & (607) & (583) & (611)& (607)& (730) \\

				\noalign{\vskip -1mm}	&[-98\%] & [-102\%] &[-147\% ]& [-125\%] &[-115\%]& [98\%] &   & [14\%] & [14\%] & [-13\%] & [10\%]& [9\%]& [117\%] &   & [-16\%] & [-16\%] & [-30\%] & [-19\%] &[-19\%]& [41\%] \\

   \noalign{\vskip 3mm}   \multicolumn{1}{l}{DW}             & -626 & -591 & -830 & -732 &-564 & 868 &   & 408 & 520 & 402 & -34& 481& 2092 &   & -246 & -176 & -264 & -495 &-223& 1136 \\
\noalign{\vskip -1mm}&      (496) & (467) & (360) & (534) & (487) & (359) &  & (691) & (588) & (426) & (845) &(672)& (471) &   & (724) & (683) & (596) & (781) &(718)& (751) \\
\noalign{\vskip -1mm}&[-71\%] & [-67\%] & [-94\%] & [-83\%]& [-64\%]& [98\%] &   &[ 23\%] & [29\%] & [22\%] & [-2\%]& [27\%]& [117\%] &   & [-9\%] & [-6\%] & [-10\%] & [-18\%]& [-8\%]& [41\%] \\

\noalign{\vskip 3mm}      \multicolumn{1}{l}{ADW}   & -597 & -599 & -1041 & -685 &-558 & 868 &   & 514 & 524 & 27 & 97 & 502& 2092 &   & -148 & -144 & -498 & -337&-165  & 1136 \\
\noalign{\vskip -1mm}                                      &        (491) & (470) & (358) & (523) & (485)& (352) &   & (663) & (582) & (428) & (793) &(653)& (458) &   & (701) & (677) & (591) & (740) & (700)& (728) \\
\noalign{\vskip -1mm}& 
     [-67\%] & [-68\%] & [-118\%] & [-77\%] & [-63\%]& [98\%]  &  & [29\%] & [29\%] & [2\%] & [5\%] & [28\%]& [117\% ]&   & [-5\%] & [-5\%] & [-18\%] & [-12\%] & [-6\%]& [41\%] \\

  \bottomrule
\end{tabular}\begin{tablenotes}[para,flushleft]
\small{
Notes: The results (standard errors are in parentheses) represent the estimated average effect of being in the experimental
sample (i.e. the estimated evaluation bias) on the 1978 earnings where the experimental control group is compared
with untreated non-experimental CPS sample. The estimated evaluation biases relative to the
experimental ATT benchmark, in percentage terms, are reported in brackets.
$\widehat{\tau}^{fe}$ is the TWFE outcome regression estimator of $\tau^{fe}$ in (\ref{fe}),  $\widehat{\tau}^{reg}$ is the OR-DID estimator (\ref{att.reg1}), $\widehat{\tau}^{dr,p}$ is the IPW DID estimator (\ref{att.ipw}), $\widehat{\tau}^{dr,p}_{std}$ is the standardized IPW DID estimator (\ref{att.ipw2}), 
$\widehat{\tau}^{dr,p}$ is our proposed DR DID estimator (\ref{dr.p.estimator}), and  $\widehat{\tau}^{dr,p}_{imp}$ is our proposed DR DID estimator (\ref{dr.p.imp.estimator}). We use a linear OR working model and a logistic PS working model, where the unknown parameters are estimated via OLS and maximum likelihood, respectively, except for $\widehat{\tau}^{dr,p}_{imp}$, where we use the estimation methods described in Section \ref{sec:firststep}.
 For each DID estimator, we report three different specifications depending on how covariates are included: ``lin.'' specification, where all
covariates enter the model linearly; ``DW'' specification, which adds to the linear
specification a dummy for zero earnings in 1974, age squared, age cubed
divided by $1000$, years of schooling squared, and an interaction term
between years of schooling and real earnings in 1974; and the \textquotedblleft ADW\textquotedblright\ specification,
which adds to the \textquotedblleft DW\textquotedblright\ specification the
interactions between married with real earnings in 1974, and between married
and zero earnings in 1974.
}
\end{tablenotes}
\end{threeparttable}
\end{adjustbox}
\label{tab:st-bias}
\end{table}
\vspace*{\fill}
\end{landscape}
}

Table \ref{tab:st-bias} highlights some interesting patterns. First,
estimators based on two-way fixed effect regression models tend to be very
stable across specifications, but usually display large positive and
statistically significant evaluation biases. Second, DID estimators based on
the regression approach tend to lead to the most precise estimates. However,
for the LaLonde sample, point estimates are severely biased downward,
leading to statistically significant evaluation biases. Abadie's IPW
estimators $\widehat{\tau }^{ipw,p}$ for the ATT tend to have the largest
standard errors across all considered estimators, but their evaluation
biases are relatively small. Like in our Monte Carlo simulation results,
considering normalized weights as in $\widehat{\tau }_{std}^{ipw,p}$ can
improve the stability of the IPW estimators $\widehat{\tau }^{ipw,p}$.
Finally, note that our proposed DR DID estimators share the favorable bias
properties of Abadie's IPW estimator, but at the same time, have smaller
standard errors than IPW estimators. When we compare $\widehat{\tau }^{dr,p}$
with $\widehat{\tau }_{imp}^{dr,p}$, we note that the further improved DR
DID estimator $\widehat{\tau }_{imp}^{dr,p}$ tends to have smaller standard
errors, particularly when one adopts the \textquotedblleft
DW\textquotedblright\ or the \textquotedblleft augmented
DW\textquotedblright\ specifications. Taken together, the results using the
NSW job training data suggest that our proposed DR DID estimators are an
attractive alternative to existing DID procedures.

\section{Concluding remarks\label{sec:conclusion}}

In this article, we proposed doubly robust estimators for the ATT in
difference-in-differences settings where the parallel trends assumption
holds only after conditioning on a vector of pre-treatment covariates. Our
proposed estimators remain consistent for the ATT when either (but not
necessarily both) a propensity score model or outcome regression models are
correctly specified, and achieve the semiparametric efficiency bound when
the working models for the nuisance functions are correctly specified. We
derived the large sample properties of the proposed estimators in situations
where either panel data or repeated cross-section data are available, and
showed that by paying particular attention to the estimation methods used to
estimate the nuisance parameters, one can form DID estimators for the ATT
that are not only DR consistent and locally semiparametric efficient, but
also DR for inference. We illustrated the attractiveness of our proposed
causal inference tools via a simulation exercise and with an empirical
application.

Our results can be extended to other situations of practical interest. A
leading case is when researchers are interested in understanding treatment
effect heterogeneity with respect to continuous covariates $X_{1}$, where $%
X_{1}$ is a (strict) subset of available covariates $X$. Here, the parameter
of interest is the conditional average treatment effect on the treated CATT$%
\left( X_{1}\right) \equiv \mathbb{E}\left[ Y\left( 1\right) -Y\left(
0\right) |X_{1},D=1\right] $ and because of its infinite dimensional nature,
the estimation and inference tools proposed in this paper are not directly
applicable. However, by combining the DR DID formulation proposed in this
paper with the methodology put forward by \cite{Chen2018a}, one can propose
uniformly valid inference procedures not only for the CATT but also for
possibly nonlinear functionals of the CATT such as (higher order) partial
derivatives, conditional average (higher order) partial derivatives, and
partial derivatives of its $log$.

Another interesting extension is when researchers want to adopt
data-adaptive, \textquotedblleft machine learning\textquotedblright\
first-step estimators instead of the parametric models discussed in this
paper. Here, the main challenge is to derive the influence function of the
DR DID estimator for the ATT, as \textquotedblleft machine
learning\textquotedblright\ estimators are, in general, in a non-Donsker
classes of functions. We envision that one can bypass such technical
complications by combining the results derived in this paper with those in 
\cite{Chernozhukov2017}, \cite{Belloni2017}, and \cite{Tan2019}, for
example; see e.g. \cite{Zimmert2019} for some recent results in this
direction. We leave the detailed analysis of these extensions to future work.

\renewcommand{\thesection}{A}

\section{Appendix A: Asymptotic Properties of the DR DID estimators based on
generic first-step estimators\label{App:conditions}}

\renewcommand{\theassumption}{A.\arabic{assumption}} \renewcommand{%
\thelemma}{A.\arabic{lemma}} \renewcommand{\theproposition}{A.%
\arabic{proposition}} \renewcommand{\thetheorem}{A.\arabic{theorem}}

\renewcommand\thetable{A.\arabic{table}} \setcounter{table}{0} %
\setcounter{lemma}{0} \setcounter{assumption}{0} \setcounter{equation}{0} %
\setcounter{figure}{0} \setcounter{proposition}{0} \setcounter{theorem}{0}

Let $g\left( x\right) $ be a generic notation for $\pi \left( x\right) $, $%
\mu _{d,t}^{p}\left( x\right) $ and $\mu _{d,t}^{rc}\left( x\right) $, $%
d,t=0,1$. Analogously and with some abuse of notation, let $g\left( x;\theta
\right) $ be a generic notation for $\pi \left( x;\gamma \right) $, $\mu
_{d,t}^{p}\left( x,\beta _{d,t}^{p}\right) $ and $\mu _{d,t}^{rc}\left(
x,\beta _{d,t}^{rc}\right) $, $d,t=0,1$. Let $W=\left(
Y_{0},Y_{1},D,X\right) $ in the panel data case and $W=\left( Y,T,D,X\right) 
$ in the repeated cross-section data case. Denote the support of $X$ by $%
\mathcal{X}$ and for a generic $Z$, let $\left\Vert Z\right\Vert =\sqrt{%
trace\left( Z^{\prime }Z\right) }$ denote the Euclidean norm of $Z$.

Let 
\begin{eqnarray*}
h^{p}\left( W;\kappa ^{p}\right) &=&\left( w_{1}^{p}\left( D\right)
-w_{0}^{p}\left( D,X;\gamma \right) \right) \left( \Delta Y-\mu _{0,\Delta
}^{p}(X;\beta _{0,0}^{p},\beta _{0,1}^{p})\right) , \\
h^{rc~1}\left( W;\kappa ^{rc~1}\right) &=&\left( w_{1}^{rc}\left( D,T\right)
-w_{0}^{rc}\left( D,T,X;\gamma \right) \right) \left( Y-\mu
_{0,Y}^{rc}\left( T,X;\beta _{0,0}^{rc},\beta _{0,1}^{rc}\right) \right) , \\
h^{rc~2}\left( W;\kappa ^{rc~2}\right) &=&\left( \left. D\right/ \mathbb{E}%
\left[ D\right] \right) \cdot \left( \mu _{1,\Delta }^{rc}\left( X;\beta
_{1,1}^{rc},\beta _{1,0}^{rc}\right) -\mu _{0,\Delta }^{rc}\left( X;\beta
_{0,1}^{rc},\beta _{0,0}^{rc}\right) \right) \\
&&+w_{1,1}^{rc}\left( D,T\right) \left( Y-\mu _{1,1}^{rc}\left( X;\beta
_{1,1}^{rc}\right) \right) -w_{1,0}^{rc}\left( D,T\right) \left( Y-\mu
_{1,0}^{rc}\left( X;\beta _{1,0}^{rc}\right) \right) \\
&&-\left( w_{0,1}^{rc}\left( D,T,X;\gamma \right) \left( Y-\mu
_{0,1}^{rc}\left( X;\beta _{0,1}^{rc}\right) \right) -w_{0,0}^{rc}\left(
D,T,X;\gamma \right) \left( Y-\mu _{0,0}^{rc}\left( X;\beta
_{0,0}^{rc}\right) \right) \right)
\end{eqnarray*}%
where $\kappa ^{p}=\left( \gamma ^{\prime },\beta _{0,0}^{p\prime },\beta
_{0,1}^{p\prime }\right) ^{\prime }$, $\kappa ^{rc~1}=\left( \gamma ^{\prime
},\beta _{0,0}^{rc\prime },\beta _{0,1}^{rc\prime }\right) ^{\prime }$ and $%
\kappa ^{rc~2}=\left( \gamma ^{\prime },\beta _{0,0}^{rc\prime },\beta
_{0,1}^{rc\prime },\beta _{1,1}^{rc\prime },\beta _{1,0}^{rc\prime }\right)
^{\prime }$. In obvious notation, the vector of pseudo-true parameter%
\footnote{%
Note that we allow for possible misspecification when we define pseudo-true
parameters.} is given by $\kappa ^{\ast ,p}$, $\kappa _{0}^{\ast ,rc~1}$,
and $\kappa ^{\ast ,rc~2}$. Let $\dot{h}^{p}\left( W;\kappa ^{p}\right)
=\partial \left. h^{p}\left( W;\kappa ^{p}\right) \right/ \partial \kappa
^{p}$ and define $\dot{h}^{rc~j}\left( W;\kappa ^{rc~j}\right) $, $j=0,1$,
analogously.

\begin{assumption}
\label{ass:parametric} $\left( i\right) $ $g\left( x\right) =g\left(
x;\theta \right) $ is a parametric model, where $\theta \in \Theta \subset 
\mathbb{R}^{k}$, $\Theta $ being compact; $\left( ii\right) $ $g\left(
X;\theta \right) $ is $a.s.$ continuous at each $\theta $ $\in \Theta $; $%
\left( iii\right) $ there exists a unique pseudo-true parameter $\theta
^{\ast }\in int\left( \Theta \right) $; $\left( iv\right) $ $g\left(
X;\theta \right) $ is $a.s.$ twice continuously differentiable in a
neighborhood of $\theta ^{\ast }$, $\Theta ^{\ast }\subset \Theta $; $\left(
v\right) $ the estimator $\widehat{\theta }$ is strongly consistent for the $%
\theta ^{\ast }$ and satisfies the following linear expansion:%
\begin{equation*}
\sqrt{n}\left( \widehat{\theta }-\theta ^{\ast }\right) =\frac{1}{\sqrt{n}}%
\sum_{i=1}^{n}l_{g}\left( W_{i};\theta ^{\ast }\right) +o_{p}\left( 1\right)
,
\end{equation*}%
where $l_{g}\left( \cdot ;\theta \right) $ is such that $\mathbb{E}\left[
l_{g}\left( W;\theta ^{\ast }\right) \right] =0$, $\mathbb{E}\left[
l_{g}\left( W;\theta ^{\ast }\right) l_{g}\left( W;\theta ^{\ast }\right)
^{\prime }\right] $ exists and is positive definite and $\lim_{\delta
\rightarrow 0}\mathbb{E}\left[ \sup_{\theta \in \Theta ^{\ast }:\left\Vert
\theta -\theta ^{\ast }\right\Vert \leq \delta }\left\Vert l_{g}\left(
W;\theta \right) -l_{g}\left( W;\theta ^{\ast }\right) \right\Vert ^{2}%
\right] =0$. In addition, $\left( vi\right) $~for some $\varepsilon >0,$ $%
0<\pi \left( X;\gamma \right) \leq 1-\varepsilon $ $a.s.$, for all $\gamma
\in int\left( \Theta ^{ps}\right) $, where $\Theta ^{ps}$ denotes the
parameter space of $\gamma .$
\end{assumption}

\begin{assumption}
\label{ass:integrabilitity} $\left( i\right) $ When panel data are
available, assume that $\mathbb{E}\left[ \left\Vert h^{p}\left( W;\kappa
^{\ast ,p}\right) \right\Vert ^{2}\right] <\infty $ and $\mathbb{E}\left[
\sup_{\kappa \in \Gamma ^{\ast ,p}}\left\vert \dot{h}^{p}\left( W;\kappa
\right) \right\vert \right] <\infty ,$ where $\Gamma ^{\ast ,p}$ is a small
neighborhood of $\kappa ^{\ast ,p}$. $\left( ii\right) $ When cross-section
data are available, assume that, for $j=1,2$, $\mathbb{E}\left[ \left\Vert
h^{rc,j}\left( W;\kappa ^{\ast ,rc,j}\right) \right\Vert ^{2}\right] <\infty 
$ and $\mathbb{E}\left[ \sup_{\kappa \in \Gamma ^{\ast ,rc~j}}\left\vert 
\dot{h}^{rc,j}\left( W;\kappa \right) \right\vert \right] <\infty ,$ where $%
\Gamma ^{\ast ,rc~~j}$ is a small neighborhood of $\kappa ^{\ast ,rc~j}$.
\end{assumption}

Assumptions \ref{ass:parametric}-\ref{ass:integrabilitity} are standard in
the literature, see e.g.~\cite{Abadie2005}, \cite{Wooldridge2007a}, \cite%
{Bonhomme2011}, \cite{Graham2012} and \cite{Callaway2018a}. Assumption \ref%
{ass:parametric} requires that the first-step estimators are based on smooth
parametric models and that the estimated parameters admit $\sqrt{n}$%
-asymptotically linear representations, whereas Assumption \ref%
{ass:integrabilitity} imposes some weak integrability conditions. Under mild
moment conditions, these requirements are fulfilled when one adopts
linear/nonlinear outcome regressions or logit/probit models, for example,
and estimates the unknown parameters by (nonlinear) least squares,
quasi-maximum likelihood, or other alternative estimation methods, see
e.g.~Chapter 5 in \cite{VanderVaart1998}, \cite{Wooldridge2007a}, \cite%
{Graham2012} and \cite{SantAnna2018a}.

Next, we derive the asymptotic properties of $\widehat{\tau }^{dr,p}$, $%
\widehat{\tau }_{1}^{dr,rc}$ and $\widehat{\tau }_{2}^{dr,rc}$ using generic
first-step estimators that satisfy Assumptions \ref{ass:parametric} and \ref%
{ass:integrabilitity}.

\subsection{Panel data case\label{sec:panel}}

In this section, we discuss the asymptotic properties of $\widehat{\tau }%
^{dr,p}$. Define $\dot{\pi}\left( x;\gamma \right) \equiv \left. \partial
\pi \left( x;\gamma \right) \right/ \partial \gamma $ and, for $t=0,1$,
define $\dot{\mu}_{0,t}^{p}\left( x;\beta _{0,t}^{p}\right) $ analogously.\
In what follows, we drop the dependence of the functionals on $W$ to ease
the notational burden. For example, we write $w_{1}^{p}=w_{1}^{p}\left(
D\right) $, $w_{0}^{p}\left( \gamma \right) =w_{0}^{p}\left( D,X;\gamma
\right) $, and so on and so forth.

For generic $\gamma $ and $\beta _{0}=\left( \beta _{0,1}^{^{\prime }},\beta
_{0,0}^{^{\prime }}\right) ^{\prime }$, let 
\begin{equation}
\eta ^{p}\left( W;\gamma ,\beta \right) =\eta _{1}^{p}\left( W;\beta
_{0}\right) -\eta _{0}^{p}\left( W;\gamma ,\beta _{0}\right) -\eta
_{est}^{p}\left( W;\gamma ,\beta _{0}\right) ,  \label{lin.rep.panel}
\end{equation}%
where 
\begin{eqnarray*}
\eta _{1}^{p}\left( W;\beta _{0}\right) &=&w_{1}^{p}\cdot \left[ \left(
\Delta Y-\mu _{0,\Delta }^{p}\left( \beta _{0}\right) \right) -\mathbb{E}%
\left[ w_{1}^{p}\cdot \left( \Delta Y-\mu _{0,\Delta }^{p}\left( \beta
_{0}\right) \right) \right] \right] , \\
\eta _{0}^{p}\left( W;\gamma ,\beta _{0}\right) &=&w_{0}^{p}\left( \gamma
\right) \cdot \left[ \left( \Delta Y-\mu _{0,\Delta }^{p}\left( \beta
_{0}\right) \right) -\mathbb{E}\left[ w_{0}^{p}\left( \gamma \right) \cdot
\left( \Delta Y-\mu _{0,\Delta }^{p}\left( \beta _{0}\right) \right) \right] %
\right] ,
\end{eqnarray*}%
and%
\begin{multline}
\eta _{est}^{p}\left( W;\gamma ,\beta _{0}\right) =l_{reg}\left( \beta
_{0}\right) ^{\prime }\cdot \mathbb{E}\left[ \left(
w_{1}^{p}-w_{0}^{p}\left( \gamma \right) \right) \cdot \dot{\mu}_{0,\Delta
}^{p}\left( \beta _{0}\right) \right]  \label{eq:est.p} \\
+l_{ps}\left( \gamma \right) ^{\prime }\cdot \mathbb{E}\left[ \alpha
_{ps}^{p}\left( \gamma \right) \left( \left( \Delta Y-\mu _{0,\Delta
}^{p}\left( \beta _{0}\right) \right) -\mathbb{E}\left[ w_{0}^{p}\left(
\gamma \right) \cdot \left( \Delta Y-\mu _{0,\Delta }^{p}\left( \beta
_{0}\right) \right) \right] \right) \cdot \dot{\pi}(\gamma )\right] ,
\end{multline}%
with $l_{reg}\left( \beta _{0}\right) =\left( l_{reg,0,1}\left( \beta
_{0,1}\right) ^{\prime },l_{reg,0,0}\left( \beta _{0,0}\right) ^{\prime
}\right) ^{\prime }$, where $l_{reg,d,t}\left( \cdot \right) $ is the
asymptotic linear representation of the estimators for the outcome
regression as described in Assumption \ref{ass:parametric}$\left( iv\right) $%
, $l_{ps}\left( \cdot \right) $ is defined analogously, $\dot{\mu}_{0,\Delta
}^{p}\left( \beta _{0}\right) =\left( \dot{\mu}_{0,1}^{p}\left( \beta
_{0,1}\right) ^{\prime },~-\dot{\mu}_{0,0}^{p}\left( \beta _{0,0}\right)
\right) ^{\prime }$ and 
\begin{equation*}
\alpha _{ps}^{p}\left( \gamma \right) =\left. \dfrac{\left( 1-D\right) }{%
\left( 1-\pi (X;\gamma )\right) ^{2}}\right/ \mathbb{E}\left[ \dfrac{\pi
\left( X;\gamma \right) \left( 1-D\right) }{1-\pi (X;\gamma )}\right] .
\end{equation*}%
For $d,t=0,1$, let $\Theta _{d,t}^{reg}\ $be the parameter space for the
regression coefficient $\beta _{d,t}$, and $\Theta ^{ps}$ be the parameter
space for the propensity score coefficient $\gamma $. Consider the following
claims:%
\begin{eqnarray}
\exists \gamma ^{\ast } &\in &\Theta ^{ps}:\mathbb{P}\left( \pi (X;\gamma
^{\ast })=p\left( X\right) \right) =1,  \label{pscore} \\
\exists \left( \beta _{0,1}^{\ast ,p},\beta _{0,0}^{\ast ,p}\right) &\in
&\Theta _{0,1}^{reg}\times \Theta _{0,0}^{reg}:\mathbb{P}\left( \mu
_{0,1}^{p}\left( X;\beta _{0,1}^{\ast ,p}\right) -\mu _{0,0}^{p}\left(
X;\beta _{0,0}^{\ast ,p}\right) =m_{0,1}^{p}\left( X\right)
-m_{0,0}^{p}\left( X\right) \right) =1.  \label{OR.p}
\end{eqnarray}

Now we are ready to state the large sample properties of $\widehat{\tau }%
^{dr,p}$.

\begin{theorem}
\label{th:att.panel} Suppose Assumptions \ref{ass:sampling}-\ref%
{ass:common.support} and Assumptions \ref{ass:parametric}-\ref%
{ass:integrabilitity} stated in Appendix \ref{App:conditions} hold.

$\left( a\right) $ Provided that either (\ref{pscore}) or (\ref{OR.p}) is
true, as $n\rightarrow \infty $, 
\begin{equation*}
\widehat{\tau }^{dr,p}\overset{p}{\rightarrow }\tau .
\end{equation*}%
Furthermore, 
\begin{eqnarray*}
\sqrt{n}(\widehat{\tau }^{dr,p}-\tau ^{dr,p}) &=&\frac{1}{\sqrt{n}}%
\sum_{i=1}^{n}\eta ^{p}\left( W_{i};\gamma ^{\ast },\beta _{0}^{\ast
,p}\right) +o_{p}(1) \\
&&\overset{d}{\rightarrow }N\left( 0,V^{p}\right) ,
\end{eqnarray*}%
where $V^{p}=\mathbb{E}[\eta ^{p}\left( W;\gamma ^{\ast },\beta _{0}^{\ast
,p}\right) ^{2}]$.

$\left( b\right) $ When both (\ref{pscore}) and (\ref{OR.p}) are true, $\eta
^{p}\left( W;\gamma ^{\ast },\beta _{0}^{\ast ,p}\right) =\eta ^{e,p}\left(
Y_{1},Y_{0},D,X\right) $ $a.s.$ and $V^{p}$ is equal to the
semiparametrically efficiency bound (\ref{eq:eff.var.p}).
\end{theorem}

Theorem \ref{th:att.panel} indicates that, provided that either the
propensity score model or the model for the evolution of the outcome for the
comparison group is correctly specified, $\widehat{\tau }^{dr,p}$ is
consistent for the ATT, implying that our proposed estimator is indeed
doubly robust. In addition, Theorem \ref{th:att.panel} indicates that our
proposed estimator admits an asymptotically linear representation and as a
consequence, it is $\sqrt{n}$-consistent and asymptotically normal. When the
models for the nuisance functions are correctly specified, our proposed DR
DID estimator is semiparametrically efficient.

Theorem \ref{th:att.panel} also suggests that one can use the analogy
principle to estimate $V^{p}$ and conduct asymptotically valid inference.%
\footnote{%
It is easy to show that the plug-in estimator of $V^{p}$ is consistent, see
e.g. Lemma 4.3 in \cite{Newey1994c} and Theorem 4.4 in \cite{Abadie2005}. We
omit the detailed derivation of this result for the sake of brevity.}
However, it is worth mentioning the fact that the exact form of $V^{p}$
depends on which nuisance models are correctly specified, implying that our
(generic) estimator $\widehat{\tau }^{dr,p}$ is doubly robust in terms of
consistency but, in general, not doubly robust for inference. Given that in
practice it is hard to know \textit{a priori} which nuisance models are
correctly specified, one should include all \textquotedblleft
correction\textquotedblright\ terms in $\eta _{est}^{p}$ when estimating $%
V^{p}$. Failing to do so may lead to asymptotically invalid inference
procedures.

\subsection{Repeated cross-section data case}

In this section, we turn our attention to our proposed DR DID estimators for
the ATT when only repeated cross-section data are available. For generic $%
\gamma $ and $\beta =\left( \beta _{1}^{^{\prime }},\beta _{0}^{^{\prime
}}\right) ^{\prime }$, where, for $d=0,1,$ $\beta _{d}=\left( \beta
_{d,1}^{^{\prime }},\beta _{d,0}^{^{\prime }}\right) ^{\prime }$, let 
\begin{equation}
\eta _{j}^{rc}\left( W;\gamma ,\beta \right) =\eta _{1}^{rc,j}\left( W;\beta
\right) -\eta _{0}^{rc,j}\left( W;\gamma ,\beta \right) -\eta
_{est}^{rc,j}\left( W;\gamma ,\beta \right) ,  \label{lin.rep.rc1}
\end{equation}%
such that, for $j=1,2$,%
\begin{eqnarray*}
\eta _{1}^{rc,j}\left( W;\beta \right) &=&\eta _{1,1}^{rc,j}\left( W;\beta
\right) -\eta _{1,0}^{rc,j}\left( W;\beta \right) , \\
\eta _{0}^{rc,j}\left( W;\gamma ,\beta \right) &=&\eta _{0,1}^{rc,j}\left(
W;\gamma ,\beta \right) -\eta _{0,0}^{rc,j}\left( W;\gamma ,\beta \right) ,
\\
\eta _{est}^{rc,j}\left( W;\gamma ,\beta \right) &=&\eta
_{est,reg}^{rc,j}\left( W;\gamma ,\beta \right) +\eta _{est,ps}^{rc,j}\left(
W;\gamma ,\beta \right) ,
\end{eqnarray*}%
and the precise definitions of all these $\eta ^{rc}$ functions are deferred
to Appendix\ \ref{App:inf.rc} to avoid excess notational complexity. An
aspect of the difference between $\eta _{1}^{rc}$ and $\eta _{2}^{rc}$ that
is worth mentioning but is perhaps buried in the notation is that $\eta
_{1}^{rc}$ depends on $\beta $ only through $\beta _{0}$, whereas $\eta
_{2}^{rc}$ depends on both $\beta _{1}$ and $\beta _{0}$. This is simply a
consequence from the fact that $\widehat{\tau }_{1}^{dr,rc}$ does not rely
on outcome regressions for the treated units, but $\widehat{\tau }%
_{2}^{dr,rc}$ does.

Consider the following claims:%
\begin{eqnarray}
\exists \left( \beta _{0,1}^{\ast ,rc},\beta _{0,0}^{\ast ,rc}\right) &\in
&\Theta _{0,1}^{reg}\times \Theta _{0,0}^{reg}:\mathbb{P}\left( \mu
_{0,1}^{rc}\left( X;\beta _{0,1}^{\ast ,rc}\right) -\mu _{0,0}^{rc}\left(
X;\beta _{0,0}^{\ast ,rc}\right) =m_{0,1}^{rc}\left( X\right)
-m_{0,0}^{rc}\left( X\right) \right) =1,  \label{OR.rc1} \\
\forall \left( d,t\right) &\in &\left\{ 0,1\right\} ^{2}\text{ }\exists
\left( \beta _{d,t}^{\ast ,rc}\right) \in \Theta _{d,t}^{reg}:\mathbb{P}%
\left( \mu _{d,t}^{rc}\left( X;\beta _{d,t}^{\ast ,rc}\right)
=m_{d,t}^{rc}\left( X\right) \right) =1.  \label{OR.rc2}
\end{eqnarray}

\begin{theorem}
\label{th:att.rc} Let $n=n_{1}+n_{0}$, where $n_{1}$ and $n_{0}$ are the
sample sizes of the post-treatment and pre-treatment periods, respectively.
Suppose Assumptions \ref{ass:sampling}-\ref{ass:common.support} and
Assumptions \ref{ass:parametric}-\ref{ass:integrabilitity} stated in
Appendix \ref{App:conditions} hold, and that $n_{1}/n\overset{p}{\rightarrow 
}\lambda \in \left( 0,1\right) $ as $n_{0},n_{1}\rightarrow \infty $.

$\left( a\right) $ Provided that either (\ref{pscore}) or (\ref{OR.rc1}) is
true, as $n\rightarrow \infty $, for $j=1,2$, 
\begin{equation*}
\widehat{\tau }_{j}^{dr,rc}\overset{p}{\rightarrow }\tau .
\end{equation*}%
Furthermore, 
\begin{eqnarray*}
\sqrt{n}(\widehat{\tau }_{j}^{dr,rc}-\tau _{j}^{dr,rc}) &=&\frac{1}{\sqrt{n}}%
\sum_{i=1}^{n}\eta _{j}^{rc}\left( W_{i};\gamma ^{\ast },\beta ^{\ast
,rc}\right) +o_{p}(1) \\
&&\overset{d}{\rightarrow }N\left( 0,V_{j}^{rc}\right) ,
\end{eqnarray*}%
where $V_{j}^{rc}=\mathbb{E}[\eta _{j}^{rc}\left( W;\gamma ^{\ast },\beta
^{\ast ,rc}\right) ^{2}]$.

$\left( b\right) $ Suppose that both (\ref{pscore}) and (\ref{OR.rc2}) are
true. Then, $\eta _{2}^{rc}\left( W;\gamma ^{\ast },\beta ^{\ast ,rc}\right)
=\eta ^{e,rc}\left( Y,D,T,X\right) $ $a.s.$, and $V_{2}^{rc}$ is equal to
the semiparametrically efficiency bound (\ref{eq:eff.var.rc}). On the other
hand, $V_{1}^{rc}\ $does not attain the semiparametric efficiency bound when
(\ref{pscore}) and (\ref{OR.rc2}) are true.
\end{theorem}

In other words, Theorem \ref{th:att.rc} states that both proposed estimators
for the ATT, $\widehat{\tau }_{1}^{dr,rc}$ and $\widehat{\tau }_{2}^{dr,rc}$%
, are doubly robust, $\sqrt{n}$-consistent and asymptotically normal.
Similar to the panel data case, the exact form of the $V_{j}^{rc}$, $j=1,2$,
depends on which working models are correctly specified, implying that the
generic estimators $\widehat{\tau }_{1}^{dr,rc}$ and $\widehat{\tau }%
_{2}^{dr,rc}$ are doubly robust in terms of consistency but in terms of
inference.

Part $\left( b\right) $ of Theorem \ref{th:att.rc} indicates that $\widehat{%
\tau }_{2}^{dr,rc}$ is semiparametrically efficient when the working model
for the propensity score, and all working models for the outcome
regressions, for both treated and comparison units, are correctly specified.
When compared to Theorem \ref{th:att.panel}$\left( b\right) $, it is evident
that such a requirement is stronger than when panel data are available.

\renewcommand{\thesection}{B}

\section{Appendix B: Influence function of the DR DID estimators with
repeated cross-section \label{App:inf.rc}}

\renewcommand{\theassumption}{B.\arabic{assumption}} \renewcommand{%
\thelemma}{B.\arabic{lemma}} \renewcommand{\theproposition}{B.%
\arabic{proposition}}

\renewcommand\thetable{B.\arabic{table}} \setcounter{table}{0} %
\setcounter{lemma}{0} \setcounter{assumption}{0} \setcounter{equation}{0} %
\setcounter{figure}{0} \setcounter{proposition}{0}

As it is evident from Theorem \ref{th:att.rc}, the influence functions of $%
\widehat{\tau }_{1}^{dr,rc}$ and $\widehat{\tau }_{2}^{dr,rc}$ play a major
role in study of the large sample properties of our proposed DR DID
estimators. In this section, we state the precise definition of $\eta
_{j}^{rc}\left( W;\gamma ,\beta \right) $, $j=1,2$, introduced in (\ref%
{lin.rep.rc1}).

We first focus on $\widehat{\tau }_{1}^{dr,rc}$. For generic $\gamma $ and $%
\beta =\left( \beta _{1}^{^{\prime }},\beta _{0}^{^{\prime }}\right)
^{\prime }$, where, for $d=0,1,$ $\beta _{d}=\left( \beta _{d,1}^{^{\prime
}},\beta _{d,0}^{^{\prime }}\right) ^{\prime }$, let%
\begin{equation*}
\eta _{1}^{rc}(W;\gamma ,\beta )=\eta _{1}^{rc,1}(W;\beta _{0})-\eta
_{0}^{rc,1}(W;\gamma ,\beta _{0})-\eta _{est}^{rc,1}(W;\gamma ,\beta _{0}),
\end{equation*}%
where 
\begin{align}
\eta _{1}^{rc,1}(W;\beta _{0})& =\eta _{1,1}^{rc,1}(W;\beta _{0,1})-\eta
_{1,0}^{rc,1}(W;\beta _{0,0}),  \label{eq:eta1.1.rc} \\
\eta _{0}^{rc,1}(W;\gamma ,\beta _{0})& =\eta _{0,1}^{rc,1}(W;\gamma ,\beta
_{0,1})-\eta _{0,0}^{rc,1}(W_{i};\gamma ,\beta _{0,0}),  \label{eq:eta1.0.rc}
\\
\eta _{est}^{rc,1}(W;\gamma ,\beta _{0})& =\eta _{est,reg}^{rc,1}(W;\gamma
,\beta _{0})+\eta _{est,ps}^{rc,1}(W;\gamma ,\beta _{0}),  \notag
\end{align}%
and, for $t=0,1$, 
\begin{align*}
\eta _{1,t}^{rc,1}(W;\gamma ,\beta )& =w_{1,t}^{rc}(D,T)\cdot \left( Y-\mu
_{0,t}^{rc}(X;\beta _{0,t})-\mathbb{E}[w_{1,t}^{rc}(D,T)\cdot \left( Y-\mu
_{0,t}^{rc}(X;\beta _{0,t})\right) ]\right) , \\
\eta _{0,t}^{rc,1}(W;\gamma ,\beta )& =w_{0,t}^{rc}(D,T,X;\gamma )\cdot
\left( Y-\mu _{0,t}^{rc}(X;\beta _{0,t})-\mathbb{E}[w_{0,t}^{rc}(D,T,X;%
\gamma )\cdot \left( Y-\mu _{0,t}^{rc}(X;\beta _{0,t})\right) ]\right) ,
\end{align*}%
and the influence functions associated with the estimation effects of the
nuisance parameters are 
\begin{equation*}
\eta _{est,reg}^{rc,1}(W;\gamma ,\beta )=l_{reg}(W;\beta )^{\prime }\cdot 
\mathbb{E}[(w_{1,1}^{rc}-w_{1,0}^{rc})-(w_{0,1}^{rc}(\gamma
)-w_{0,0}^{rc}(\gamma ))\cdot \dot{\mu}_{0,Y}^{rc}(T,X;\beta )],
\end{equation*}%
and 
\begin{multline}
\eta _{est,ps}^{rc,1}(W;\gamma ,\beta )  \notag \\
=l_{ps}(D,X;\gamma )^{\prime }\cdot \mathbb{E}\left[ \alpha
_{ps,1}^{rc}(\gamma )\cdot \left( Y-\mu _{0,1}^{rc}(X;\beta _{0,1})-\mathbb{E%
}[w_{0,1}^{rc}(\gamma )\cdot \left( Y-\mu _{0,1}^{rc}(\beta _{0,1})\right)
]\right) \dot{\pi}(X;\gamma )\right] \\
-l_{ps}(D,X;\gamma )^{\prime }\cdot \mathbb{E}\left[ \alpha
_{ps,0}^{rc}(\gamma )\cdot \left( Y-\mu _{0,0}^{rc}(\beta _{0,0})-\mathbb{E}%
[w_{0,0}^{rc}(\gamma )\cdot \left( Y-\mu _{0,0}^{rc}(\beta _{0,0})\right)
]\right) \dot{\pi}(X;\gamma )\right] ,  \notag
\end{multline}%
where, for $t=0,1$,%
\begin{equation*}
\alpha _{ps,t}^{rc}(\gamma )\equiv \alpha _{ps,t}^{rc}(D,T,X;\gamma )=\left. 
\frac{(1-D)1\left\{ T=t\right\} }{(1-\pi (X;\gamma ))^{2}}\right/ \mathbb{E}%
\left[ \frac{\pi (X;\gamma )(1-D)1\left\{ T=t\right\} }{1-\pi (X;\gamma )}%
\right] ,
\end{equation*}%
and $w_{1,t}^{rc}\equiv w_{1,t}^{rc}\left( D,T\right) $, $%
w_{0,t}^{rc}(\gamma )\equiv w_{0,t}^{rc}(D,T,X;\gamma )$.

The influence function of $\widehat{\tau }_{2}^{dr,rc}$ is given by%
\begin{equation*}
\eta _{2}^{rc}(W;\gamma ,\beta )=\eta _{1}^{rc,2}(W;\beta )-\eta
_{0}^{rc,2}(W;\gamma ,\beta _{0})-\eta _{est}^{rc,2}(W;\gamma ,\beta _{0}),
\end{equation*}%
where 
\begin{align}
\eta _{1}^{rc,2}(W;\beta )& =\eta _{1,1}^{rc,2}(W;\beta )-\eta
_{1,0}^{rc,2}(W;\beta ),  \label{eq:eta.2.1.rc} \\
\eta _{0}^{rc,2}(W;\gamma ,\beta _{0})& =\eta _{0}^{rc,1}(W;\gamma ,\beta
_{0}),  \label{eq:eta.2.0.rc} \\
\eta _{est}^{rc,1}(W;\gamma ,\beta _{0})& =\eta _{est}^{rc,1}(W;\gamma
,\beta _{0}),  \notag
\end{align}%
and, for $d=0,1$, $\mu _{d,\Delta }^{rc}\left( X;\beta _{d,1},\beta
_{d,0}\right) \equiv \mu _{d,1}^{rc}\left( X;\beta _{d,1}\right) -\mu
_{d,0}^{rc}\left( X;\beta _{d,0}\right) $, and%
\begin{eqnarray*}
\eta _{1,1}^{rc,2}(W;\beta ^{\ast }) &=&\frac{D}{\mathbb{E}\left[ D\right] }%
\left( \mu _{1,\Delta }^{rc}\left( X;\beta _{1,1},\beta _{1,0}\right) -%
\mathbb{E}\left[ \frac{D}{\mathbb{E}\left[ D\right] }\mu _{1,\Delta
}^{rc}\left( X;\beta _{1,1},\beta _{1,0}\right) \right] \right) \\
&&+w_{1,1}^{rc}(D,T)\cdot \left( \left( Y-\mu _{1,1}^{rc}\left( X;\beta
_{1,1}\right) \right) -\mathbb{E}[w_{1,1}^{rc}\cdot \left( Y-\mu
_{1,1}^{rc}\left( X;\beta _{1,1}\right) \right) ]\right) , \\
\eta _{1,0}^{rc,2}(W;\beta ) &=&\frac{D}{\mathbb{E}\left[ D\right] }\left(
\mu _{0,\Delta }^{rc}\left( X;\beta _{0,1}^{rc},\beta _{0,0}^{rc}\right) -%
\mathbb{E}\left[ \frac{D}{\mathbb{E}\left[ D\right] }\mu _{0,\Delta
}^{rc}\left( X;\beta _{0,1},\beta _{0,0}\right) \right] \right) \\
&&+w_{1,0}^{rc}(D,T)\cdot \left( Y-\mu _{1,0}^{rc}\left( X;\beta
_{1,0}\right) -\mathbb{E}[w_{1,0}^{rc}\cdot \left( Y-\mu _{1,0}^{rc}\left(
X;\beta _{1,0}\right) \right) ]\right) .
\end{eqnarray*}%
Note that estimating the OR coefficients associated with the treated group
does not lead to any estimation effect.

\renewcommand{\thesection}{C}

\onehalfspacing{\small
\bibliographystyle{jasa}
\bibliography{DID}
}

\end{document}